\documentclass[aps,pre,showpacs,twocolumn]{revtex4}
\newcommand{\be}{\begin{eqnarray}}
\newcommand{\ee}{\end{eqnarray}}
\newcommand{\bw}{\begin{widetext}}
\newcommand{\ew}{\end{widetext}}
\newcommand{\ba}{\begin{array}}
\newcommand{\ea}{\end{array}}
\newcommand{\no}{\nonumber}
\newcommand{\tr}{\mbox{tr}\,}
\newcommand{\str}{\mbox{str}\,}
\newcommand{\e}{\epsilon}
\newcommand{\Si}{\Sigma}

\begin{document}
\title{Nonlinear $\sigma$ model approach for level correlations 
in chiral disordered systems}

\author{Kazutaka Takahashi\footnote{Present address: RIKEN, Wako, Saitama 351-0198, Japan}}
\affiliation{Theoretische Physik III, 
 Ruhr-Universit\"at Bochum, 44780 Bochum, Germany}
\date{\today}

\begin{abstract}
 We study level correlations of 
 disordered systems with chiral unitary symmetry (AIII symmetry).
 We use a random matrix model with a finite correlation length 
 to derive a supersymmetric nonlinear $\sigma$ model.
 The result is compared with existing results based on other models.
 Using the methods by Kravtsov and Mirlin 
 (Pis'ma Zh. Eksp. Teor. Fiz. {\bf 60}, 645 (1994) 
 [JETP Lett. {\bf 60}, 656 (1994)]) 
 and Andreev and Altshuler [Phys. Rev. Lett. {\bf 75}, 902 (1995)],
 we calculate the density of states and two-level correlation function.
 The result is expressed using the spectral determinant 
 as in traditional nonchiral systems.
 We discuss the renormalization of the mean level spacing
 which is not present in the traditional systems.
\end{abstract}
\maketitle

%%%%%%%%%%%%%%%%%%%%%%%%%%%%%%%%%%%%%%%%%%%%%%%%%%%%%%%%%%%%%%%%%%%%%%%%%%%%
%%%%%%%%%%%%%%%%%%%%%%%%%%%%%%%%%%%%%%%%%%%%%%%%%%%%%%%%%%%%%%%%%%%%%%%%%%%%
\section{Introduction}

% universality, chiral symmetry

 The classification of disordered systems
 is based on symmetries of the Hamiltonian.
 According to invariance properties 
 under time-reversal and spin rotation, 
 three symmetry classes -- unitary, orthogonal, and symplectic -- 
 are well known since the work by Wigner and Dyson \cite{WD}.
 The modern classification is based on 
 the notion of symmetric spaces \cite{Z} and 
 indicates that ten universality classes exist.
 Although there was an early effort at 
 a universality classification in 80's \cite{Hikami}, 
 the additional seven classes did not attract 
 much attention until physical applications were found \cite{ChRMT,AZ}.
 The importance of chiral symmetry in disordered systems 
 was first noticed in Ref.\cite{ChRMT} 
 by using random matrix theory (RMT) in the context of 
 quantum chromo dynamics (QCD) and mesoscopic quantum wires.
 In systems with chiral symmetry,
 eigenvalues of the Hamiltonian appear in pairs $\pm \e$
 and the origin $\e=0$ plays a special role for level correlations.

% method, role of zero and nonzero modes

 In order to analyze such systems, 
 the supersymmetry method \cite{Efetov} is known to be a useful tool
 for both perturbative and nonperturbative calculations. 
 This method allows one to obtain a nonlinear $\sigma$ model
 with supermatrix fields as effective modes.
 One can discuss weak localization effects using perturbation theory, 
 where an expansion in terms of 
 diffusion propagators (\ref{pi}) is performed.
 A diagrammatical interpretation is thus possible, and 
 weak localization implies a large conductance $g\gg 1$,
 where $g$ is proportional to the diffusion constant in the propagator.
 The localization property can also be discussed 
 using the renormalization group method. 
 This expansion is justified only for nonzero modes 
 $q\ne 0$ in the propagator.
 The zero mode sector contains a totally different contribution
 and gives the ergodic result $g=\infty$.
 Using the zero mode, we can calculate level correlation functions
 scaled in terms of the mean level spacing \cite{Efetov}. 
 The result is nonperturbative, parameter-free, and universal. 
 We know that treating the zero mode perturbatively
 gives only the asymptotic form of the exact result.

% calculation beyond the universality

 Thus it is important to notice the different roles of 
 the zero and nonzero modes.
 At finite $g$, the nonzero modes modify the universal result 
 of level correlation functions \cite{AS,KM,AA}.
 Kravtsov and Mirlin (KM) treated the zero and nonzero modes 
 separately and found finite-$g$ corrections 
 to the universal result \cite{KM}.
 Due to technical problems, the result was restricted to the domain 
 $z\ll g$ where $z$ is the scaled energy variable.
 Using another method, Andreev and Altshuler (AA) 
 considered the domain $z\gg 1$ 
 where the perturbative expansion makes sense \cite{AA}.
 They reached the nonperturbative regime 
 by noticing the existence of a set of nontrivial saddle points.
 Considering the expansion around two saddle points 
 the result was expressed using 
 the determinant of the diffusion propagator, 
 which is called the spectral determinant in the literature.
 Although their method did not treat the zero and nonzero modes separately, 
 it was shown in Ref.\cite{MI} that 
 the separation, just as in KM's method, gives the same result.

 Using the derived result, the authors in Ref.\cite{AA} 
 found a smearing of the singularity at the Heisenberg time
 in the form factor (the Fourier transform of 
 the two-level correlation function).
 Furthermore, the use of the spectral determinant represents a link 
 from disordered to chaotic systems.
 The authors in Ref.\cite{AAA} noticed that a similar treatment 
 can be applied to general chaotic systems 
 just by replacing the diffusion operator in the spectral determinant 
 by the Perron-Frobenius operator.
 For a chaotic system, the expression of the determinant using 
 the trace formula was discussed in Ref.\cite{BK}.
 Thus the expression using the spectral determinant is 
 important for a unified treatment of disordered and chaotic systems. 
 The result was applied to critical statistics \cite{KT} and  
 the relation to the density-density correlation in the Calogero-Sutherland 
 model at finite temperature was discussed.

% our aim

 In this paper we consider systems with chiral unitary symmetry.
 Starting from a chiral random matrix model 
 with a finite correlation length, 
 we derive a nonlinear $\sigma$ model 
 and calculate the density of states (DOS) and 
 two-level correlation function (TLCF).
 Our aim in this paper is not to discuss a specific model
 but to discuss the generic properties of chiral symmetric systems.
 Actually the $\sigma$ model we use in this paper is believed to be 
 applicable to a broad range of physical systems and 
 we discuss the relation to other $\sigma$ models for specific systems.
 Then, we calculate the DOS and TLCF using a nonperturbative method
 which is equivalent to both methods by KM and AA.
 Out method is similar to that in Ref.\cite{MI} and 
 the zero and nonzero modes are separated explicitly.
 For chiral symmetric systems, 
 the calculation using the KM method has been carried out in Ref.\cite{TI}.
 In contrast to the approach in Refs.\cite{KM,TI}, 
 we integrate the zero mode first, and then, 
 treat the nonzero modes perturbatively.
 The advantage of this method is that
 all domains are treated in a unified way.
 We also discuss the effect of the DOS renormalization, 
 which is specific for nonstandard symmetry systems.
 We restrict our discussion to chiral unitary symmetry 
 (AIII symmetry) and the extentions to other chiral symmetric classes, 
 chiral orthogonal (BDI) and chiral symplectic (CII), 
 will be discussed elsewhere.

% organization of this paper

 The organization of this paper is as follows.
 In Sec.\ref{sigma}, starting from the random Hamiltonian,  
 we derive the supersymmetric nonlinear $\sigma$ model.
 It differs from the traditional $\sigma$ model 
 written in terms of a supermatrix $Q$ by symmetries of the matrix  
 and the presence of an additional term.
 We discuss relations to other models.
 Next, the DOS and TLCF are calculated in Sec.\ref{level}.
 In Sec.\ref{double}, we discuss the effect of the additional term 
 and the DOS renormalization.
 Sec.\ref{conc} is devoted to discussions and conclusions.

%%%%%%%%%%%%%%%%%%%%%%%%%%%%%%%%%%%%%%%%%%%%%%%%%%%%%%%%%%%%%%%%%%%%%%%%%%%%
%%%%%%%%%%%%%%%%%%%%%%%%%%%%%%%%%%%%%%%%%%%%%%%%%%%%%%%%%%%%%%%%%%%%%%%%%%%%
\section{Supersymmetric nonlinear $\sigma$ model}
\label{sigma}

%%%%%%%%%%%%%%%%%%%%%%%%%%%%%%%%%%%%%%%%%%%%%%%%%%%%%%%%%%%%%%%%%%%%%%%%%%%%
\subsection{Derivation}

 In this paper we treat the Hamiltonian in the form 
\be
 H = \left(\ba{cc} 0 & W \\ W^\dag & 0 \ea\right),
\ee
 where $W$ is an arbitrary matrix.
 This Hamiltonian possesses chiral symmetry, 
 which means that the eigenvalues appear in pairs $\pm \e_i$.
 The matrix $W$ can be a rectangular matrix ($n\times m$)
 as well as a square one ($n=m$).
 $\nu=|n-m|$ is the topological number
 and is equal to the number of zero eigenvalues of the Hamiltonian.
 Here we consider $\nu=0$ and 
 the extention to a finite $\nu$ will be discussed elsewhere.
 Making a unitary transformation, we have 
\be
 H = \left(\ba{cc} \Omega_1 & \Omega_2 \\ 
 \Omega_2 & -\Omega_1 \ea\right). \label{H}
\ee
 $\Omega_{1,2}$ are $n\times n$ Hermitian matrices. 
 Treating these matrices as random ones, 
 we can obtain the original chiral RMT \cite{ChRMT}.
 Due to the chiral structure of the Hamiltonian,
 two random matrices couple in the single Hamiltonian 
 and nontrivial correlations of the single Green function are expected.
 We restrict our discussion to the chiral unitary ensemble, 
 which means $\Omega_{1,2}$ are arbitrary Hermitian matrices.

 We consider a system written in field theoretical form as 
\be
 {\cal H} = \int_{xy} \psi^\dag(x) H(x,y)\psi(y),
\ee
 where $\psi$ is the fermionic field operator and $\int_x=\int d^dx$.
 The random Hamiltonian $H(x,y)$ has the chiral structure (\ref{H}) and 
\be
 \Omega_{1,2}(x,y) = \omega_{1,2}(x,y)a(|x-y|).
\ee
 $\omega_{1,2}$ are random matrices and are averaged 
 using the Gaussian integral
\be
 \left<\cdots\right> &=& 
 \int{\cal D}\omega_{1,2}\left(\cdots\right)
 \exp\biggl[-\frac{1}{\lambda^2}
 \int_{xy}(\omega_1(x,y)\omega_1(y,x) \no\\
 & & +\omega_2(x,y)\omega_2(y,x))\biggr], 
\ee
 where $\lambda$ is a free parameter.
 The function $a(r)$ represents
 a finite correlation of the Hamiltonian.
 We assume the range of the correlation, denoted by $r_0$, is large 
 so that the saddle point approximation 
 is applicable in the following calculation.
 In the limit $r\ll r_0$, $a(r)\sim 1$
 and we have the fully Gaussian correlation.
 In the opposite limit $r\gg r_0$, 
 we assume the correlation decays fast enough,
 e.g., $a(r)\sim \exp(-|r|/r_0)$.

 This finite-range model is more realistic than chiral RMT
 in which all the matrix elements correlate with each other in the same way.
 The finite-range effect can be realized 
 as a weak localization correction and 
 a new energy scale $E_c=D/L^2$ (Thouless energy),  
 where $D$ is the diffusion constant and $L$ the system length,
 comes into the analysis.
 Another interesting situation is when the decay of the matrix is power-law. 
 For a certain range of parameters
 this model reproduces the physics of the Anderson transition \cite{PRBM}. 
 Extentions of the present work to the power-law case 
 are discussed in Ref.\cite{GT}.

 We mention related works \cite{Gade,th,rfm,GWW,chpt} 
 in which similar nonlinear $\sigma$ models were considered 
 for systems with chiral symmetry. 
 Our model is a simple generalization of models used in \cite{Gade,th}.
 In other works, 
 the random flux model \cite{rfm},
 the random gauge field model \cite{GWW}, and 
 the partially quenched chiral perturbation theory 
 as the low-energy model of QCD \cite{chpt}
 were considered.
 The derived nonlinear $\sigma$ models differ from 
 the standard diffusion model for nonchiral systems
 by symmetries of the matrix.
 Furthermore, an additional term was found in Refs.\cite{Gade,rfm,chpt}
 although it was not found in Refs.\cite{th,GWW}.
 Here we rederive the $\sigma$ model and 
 discuss relations to these models.
 In fact the additional term can exist and can be derived 
 by a careful treatment of the massive modes integration.
 Although these models are different, 
 we expect common low-energy properties. 
 Our goal is to investigate them in the framework of 
 the nonlinear $\sigma$ model.

 Let us derive the nonlinear $\sigma$ model 
 using the supersymmetry method.
 Our derivation is similar to 
 that in Refs.\cite{RBM,PRBM}.
 We first define the generating function for the single Green function.
 Following the Efetov's notation and conventions \cite{Efetov}, 
 we define it as $Z_1[J]=\int {\cal D}(\psi,\bar{\psi})\exp(-{\cal L})$, 
 with 
\be
 {\cal L} &=& -i\int_{xy}\bar{\psi}(x)[\e^{+}\delta(x-y) \no\\
 & & -H(x,y)+kJ(x)\delta(x-y)]\psi(y),
\ee
 where $k=\mbox{diag}(1,-1)$ operates in superspace, 
 $\psi$ is a four-component supervector and $\bar{\psi}=\psi^\dag$.
 The source field $J$ is a $2\times 2$ matrix in chiral space.
 We take the ensemble averaging to obtain 
\be
 {\cal L}
 &=& -\frac{1}{4}\int_{xy}A(x,y)\str \tilde{\rho}(x)\tilde{\rho}(y) \no\\
 & & -i\int_{x}\bar{\psi}(x)\left[\e^{+}+kJ(x)\right]\psi(x),
\ee
 where $A(x,y)=a^2(x-y)$, and
\be
 \tilde{\rho}(x) &=& \frac{1}{\sqrt{2}}\left[
 \rho(x)-\Si_x\rho(x)\Si_x\right], \no\\
 \rho(x) &=& \Si_z^{1/2}\psi(x)\bar{\psi}(x)\Si_z^{1/2}.
\ee
 $\Si_{x,z}$ are the Pauli matrices in chiral space.
 The Hubbard-Stratonovich field $Q$ is introduced in the standard way.
 After integrations over $\psi$ and $\bar{\psi}$,  
 we have $\left<Z_1[J]\right>=\int {\cal D}Q \exp(-F_1[J])$ with 
\be
 F_1[J]
 &=& \frac{A_0}{2}\int_{xy}(A^{-1})(x,y)\str Q(x)Q(y) \no\\
 & & -\str\ln \left(\e^{+}\Si_z+Jk\Si_z+i\lambda\sqrt{A_0}Q\right),
\ee
 where $A_0=\int_y A(x,y)\sim r_0^d$.
 $Q$ is a 4$\times$4 supermatrix and 
 has the same symmetry as $\tilde{\rho}(x)$, 
 which gives the condition $\{Q,\Si_x\}=0$.

 We consider the saddle-point approximation.
 We are interested in the vicinity of the origin $\e=0$
 where chiral symmetry becomes important.
 At this point, the saddle point equation gives $Q^2=1$ and 
 the saddle-point manifold is obtained as $Q=T\Si_z\bar{T}$ 
 where $\bar{T}$ is the inverse of $T$.
 Symmetries of the $T$-matrix were considered in Ref.\cite{AST} 
 and the explicit parametrization was obtained as
\be
 T = \sqrt{1-P^2}-iP, \quad
 P = \left(\ba{cc} 0 & t \\ t & 0 \ea\right), \quad
 t = \left(\ba{cc} a & \sigma \\ \rho & ib \ea\right), \no\\\label{P0}
\ee
 where $a$, $b$ are real variables and 
 $\sigma$, $\rho$ grassmann variables.
 In addition, we must take into account 
 the massive degrees of freedom which are not on the saddle point manifold.
 Usually, in nonchiral systems, integrations of 
 the massive degrees of freedom do not give any contribution.
 However, in the present case, the integrations give 
 additional contributions written in terms of the massless modes.
 We can write the $Q$-matrix as $Q=T(\Si_z+\delta Q)\bar{T}$ 
 where $\delta Q$ denotes the massive modes and changes the saddle point.
 Since the $Q$-matrix anticommutes with $\Si_x$, 
 the structure of $\delta Q$ in chiral space is determined as 
\be
 \delta Q = \left(\ba{cc} \delta q & 0 \\ 0 & -\delta q \ea\right),
 \label{deltaQ}
\ee
 where $\delta q$ is a 2$\times$2 supermatrix.
 This $Q$ is substituted to the generating function and 
 the functional $F_1$ is expanded in powers of $\delta Q$.
 We have 
\be
 F_1[J] = F_1^{(0)}[J]+\tilde{F}_1^{(0)}+F_1^{({\rm I})},
\ee
 where 
\be
 F_1^{(0)}[J] &=& 
 \frac{A_0}{2}\int_{xy}R(x,y)\str Q(x)Q(y) \no\\
 & & +\frac{i\pi\e}{2V\Delta}\int_x\str \Si_zQ(x) \no\\
 & & +\frac{i\pi}{2V\Delta}\int_{x}\str J(x)k\Si_zQ(x), \no\\
 \tilde{F}_1^{(0)} &=& 
 \frac{A_0}{2}\int_{xy} \left[(A^{-1})(x,y)
 +\delta(x-y)A_0^{-1}\right] \no\\
 & & \times 
 \str \delta Q(x)\delta Q(y), \no\\
 F_1^{({\rm I})} &=& 
 \frac{A_0}{2}\int_{xy}R(x,y) \no\\
 & & \times\str\bigl\{
 2\left[\bar{T}(y) Q(x)T(y)-\Si_z\right]\delta Q(y) \no\\
 & & +T(x)\delta Q(x)\bar{T}(x) T(y)\delta Q(y)\bar{T}(y) \no\\
 & & -\delta Q(x)\delta Q(y) +\cdots\bigr\}.
\ee
 $Q(x)=T(x)\Si_z\bar{T}(x)$, 
 $R(x,y) = A^{-1}(x,y)-\delta(x-y)A_0^{-1}$, and 
 $\Delta=\pi\lambda\sqrt{A_0}/2V$ ($V$ is the system volume, and 
 we put the lattice constant $a=1$)
 is the inverse of the DOS (mean level spacing) at $\e=0$.
 $F_1^{(0)}[J]$ is independent of the massive modes, 
 $\tilde{F}_1^{(0)}$ is the purely massive mode, 
 and $F_1^{({\rm I})}$ is the mixing term.
 Using the cumulant expansion and 
 integrations of the massive modes we obtain 
 $F_1 \sim F_1^{(0)}[J]+\langle F_1^{({\rm I})}\rangle_{\tilde{F}_1^{(0)}}$
 where
\be
 \langle F_1^{({\rm I})}\rangle_{\tilde{F}_1^{(0)}} 
 &=& \frac{1}{4}\int_{xy}R(x,y) \no\\
 & & \times
 [\str \bar{T}(y)T(x)  \str\bar{T}(x)T(y) \no\\
 & & -\str \bar{T}(y)T(x)\Si_x  \str\bar{T}(x)T(y)\Si_x].
\ee
 This calculation can be systematically done by using 
 contraction rules derived in Appendix \ref{cont}.
 We neglected contributions that can be considered higher order ones.
 The first term in the above equation is also neglected 
 since the expansion does not include second order in $P$ 
 [see Eq.(\ref{P0})].
 We obtain 
\be
 F_1 &=&  \frac{A_0}{2}\int_{xy}R(x,y)\str Q(x)Q(y) \no\\
 & & -\frac{1}{4}\int_{xy}R(x,y)
 \str \bar{T}(y)T(x)\Si_x  \str\bar{T}(x)T(y)\Si_x \no\\
 & & +\frac{i\pi\e}{2\Delta V}\int_x\str \Si_zQ(x) \no\\
 & & +\frac{i\pi}{2\Delta V}\int_{x}\str J(x)k\Si_zQ(x). \label{F0}
\ee
 The second term has a double-supertrace form 
 and is not present in nonchiral systems.
 The crucial point is that 
 the massive modes were parametrized as in Eq.(\ref{deltaQ}).
 They have the structure $\Si_z$ in chiral space.
 $\delta Q$ in a form $\delta Q=\mbox{diag}(\delta q_1, \delta q_2)$
 would give the first term of $\langle F_1^{(I)}\rangle$ only, 
 which is the case for nonchiral systems.

 Using the gradient expansion,
 we obtain the final form of the $\sigma$ model 
\be
 F_1 &=& \frac{\pi D}{4\Delta V}\int\str (\nabla Q)^2
 -\frac{\pi D_1}{16\Delta V}\int\left(\str Q\nabla Q\Si_x\right)^2 \no\\
 & & +\frac{i\pi\e}{2\Delta V}\int\str Q\Si_z, \label{F1}
\ee
 where we neglected the source term, 
 $Q(x)=T(x)\Si_z\bar{T}(x)$ is a $4\times 4$ supermatrix, and 
\be
 \frac{\pi D}{\Delta V} 
 = \frac{\int_r r^2 a^2(r)}{\int_r a^2(r)}, \quad
 \frac{\pi D_1}{\Delta V} 
 = \frac{\int_r r^2 a^2(r)}{\left[\int_r a^2(r)\right]^2}.
\ee
 Due to the relation $D=D_1\int_r a^2(r)\sim D_1r_0^d$, 
 the constant $D_1$ is smaller than $D$ by the factor $1/r_0^d$
 and the second term in Eq.(\ref{F1}) can be neglected.
 However, it can be important when the quantum effect is 
 taken into account by the renormalization group method. 
 It is discussed in Sec.\ref{double}.

 The generating function $Z_1$ is used only for a single Green function.
 It is straightforward to extend the calculation to the case 
 of products of Green functions.
 The generating function for the product of the retarded Green function 
 $\left<\tr G^{(R)}(\e_1)\tr G^{(R)}(\e_2)\right>$ is defined as
\be
 Z_2[J]
 = \int {\cal D}(\psi,\bar{\psi})\exp
 \left[i\int\bar{\psi}(\hat{\e}^{+}-H+kJ)\psi\right], 
\ee
 where $\psi$, $\bar{\psi}$ are eight-component supervectors.
 $\hat{\e}=\mbox{diag}(\e_1,\e_2)$ is the matrix in ``two-point'' space.
 In chiral symmetric systems, the identity 
 $\mbox{tr} G^{(A)}(\e) = -\mbox{tr} G^{(R)}(-\e)$ 
 holds and the generating function for the advanced Green function 
 can be found from $Z_2$.
 Repeating the calculation in a similar way, 
 we find the $\sigma$ model 
 $\left<Z_2\right> = \int{\cal D}Q\exp(-F_2)$ with 
\be
 F_2 &=& \frac{\pi D}{4\Delta V}\int\str (\nabla Q)^2 \no\\
 & & -\frac{\pi D_1}{32\Delta V}\int\left[
 \left(\str Q\nabla Q\Si_x\right)^2
 +\left(\str \Lambda Q\nabla Q\Si_x\right)^2\right] \no\\
 & & +\frac{i\pi}{2\Delta V}\int\str \hat{\e}\Si_z Q,
 \label{F2}
\ee
 where $Q=T\Si_z\bar{T}$ is an $8\times 8$ supermatrix
 and $\Lambda=\mbox{diag}(1,-1)$ in two-point space.

%%%%%%%%%%%%%%%%%%%%%%%%%%%%%%%%%%%%%%%%%%%%%%%%%%%%%%%%%%%%%%%%%%%%%%%%%
\subsection{Comparison with other models}

 Our derived nonlinear $\sigma$ model  
 is equivalent to the models in Refs.\cite{TI,th}
 except for the presence of the double trace term.
 The reason why that term was absent in Refs.\cite{TI,th}
 is that the massive mode integration was not taken carefully.

 In order to compare our result with the models in Refs.\cite{Gade,rfm}
 we use the $Q$-matrix parametrization  
\be
 Q = \Si_z \bigl[(1-P^2)^{1/2}+iP\bigr]^2, \quad
 P = \left(\ba{cc} 0 & t \\ t & 0 \ea\right), 
\ee
 where $t$ is a 2$\times$2 supermatrix.
 The random flux model in Ref.\cite{rfm}
 is mapped onto the effective action 
\be
 S_{RF} &=& -\frac{2}{b}\int \str \nabla T^{-1}\nabla T
 -\frac{1}{c}\int \left(\str T^{-1}\nabla T\right)^2 \no\\
 & & -\frac{2i\omega}{b}\int \str \left(T+T^{-1}\right),
 \label{as}
\ee
 where $T\in\mbox{GL}(n|n)$.
 This model is reduced to our model by using the parametrization 
\be
 T = [t+(1+t^2)^{1/2}]^2,
\ee
 and putting $n=1$.
 The ``flavor'' degrees of freedom $n$
 represents different species of electrons and 
 are not important for the present problem.
 We note that different notation and conventions are used 
 in this expression.
 In contrast with our definition of supermathematics \cite{Efetov}, 
 the definition in Ref.\cite{VWZ} was used in Eq.(\ref{as}),
 which explains the difference in appearance  
 between Eqs.(\ref{F1}) and (\ref{as}).
 
 It is worthwhile to mention 
 the relation of the coupling constants $b$ and $c$.
 The authors in Ref.\cite{rfm} found the relation $b\sim c/N$ 
 where $N$ are the ``color'' degrees of freedom.
 $N$ must be large in order to justify the saddle point approximation.
 Thus the second term in Eq.(\ref{as}),
 is small compared with the first term.
 This is precisely what we found and 
 the correlation length $r_0$ corresponds to $N$.
 We also note that we neglected 
 the topological term coming from the boundary condition \cite{rfm}.
 Such a term is expected to be derived in our model
 by considering a finite topological number $\nu$  
 and it will be discussed elsewhere.

 In a similar way, our result is compared with the Gade's 
 replica $\sigma$ model based on the sublattice models \cite{Gade} 
\be
 S_G &=& \frac{2}{b}\int\tr \nabla (Z+W)\nabla (Z-W) \no\\
 & & -\frac{1}{c}\int \left[\tr (W\nabla Z-Z\nabla W) \right]^2 
 -\frac{4i\omega}{b}\int \tr W, \no\\
\ee
 where $Z$ is a matrix with some symmetry and $W=(1+Z^2)^{1/2}$.
 The parametrization
\be
 Z = 2t(1+t^2)^{1/2}, \quad W = 1+2t^2,
\ee
 is used to find a formal agreement with our model.
 We note that the Gade's model was obtained by using the replica method 
 and the structure of the matrix $t$ is different from ours.
 However we show in the following that, 
 at least in the perturbative regime, 
 both calculations give the same result.
 It is known that the replica and supersymmetry methods 
 give the same perturbative result for the same symmetry class.

 The relation of the coupling constants $b$ and $c$ 
 was not discussed in Ref.\cite{Gade}.
 It is not clear what is the large parameter in the model 
 to justify the saddle-point approximation. 
 It is expected that introduction of such a parameter 
 leads to a similar relation just as in other calculations.

 Both works \cite{rfm,Gade} did not use the $Q$-matrix representation.
 It has been used in traditional $\sigma$ models 
 and is useful for comparison of models and 
 for formulation of perturbative and 
 nonperturbative calculations as we demonstrate below.
 It is also important to find gauge invariance of the model.
  
%%%%%%%%%%%%%%%%%%%%%%%%%%%%%%%%%%%%%%%%%%%%%%%%%%%%%%%%%%%%%%%%%%%%%%%%%%%%
%%%%%%%%%%%%%%%%%%%%%%%%%%%%%%%%%%%%%%%%%%%%%%%%%%%%%%%%%%%%%%%%%%%%%%%%%%%%
\section{Level correlation functions}
\label{level}

 In this section, we calculate the DOS and TLCF by using 
 the nonlinear $\sigma$ models derived in the previous section.
 We neglect the double-trace term contribution and put $D_1=0$.
 This is because $D_1$ is smaller than $D$ by the factor $1/r_0$
 at the classical level.
 The effect of the double-trace term is discussed in Sec.\ref{double}.

 We write down the DOS and TLCF in a functional integral form.
 The DOS is given by 
\be
 \left<\rho(\e)\right> 
 = \frac{1}{4\Delta V}\mbox{Re}\int {\cal D}Q
 \left[\int_x \str k\Si_z Q(x)\right] \mbox{e}^{-F_1}, \label{rho}
\ee
 where $F_1$ is given by Eq.(\ref{F1}) (we put $D_1=0$)
 and $Q$ is a 4$\times$4 supermatrix.
 The TLCF is
\be
 \left<\rho(\e_1)\rho(\e_2)\right> &=& 
 \frac{1}{4}[
 W(\e_1,\e_2)+W(\e_1,-\e_2) \no\\
 & & +W(-\e_1,\e_2)+W(-\e_1,-\e_2)], \label{TLCF}\\
 W(\e_1,\e_2)
 &=& \frac{1}{16\Delta^2V^2}\int {\cal D}Q
 \left[\int_x \str k\Lambda_1\Si_z Q(x)\right] \no\\
 & & \times\left[\int_y \str k\Lambda_2\Si_z Q(y)\right] 
 \mbox{e}^{-F_2}, \label{W}
\ee
 where $F_2$ is given by Eq.(\ref{F2}), 
 $Q$ is an 8$\times$8 supermatrix, and 
 $\Lambda_{1,2}=(1\pm\Lambda)/2$.
 In the following we use the connected part of the TLCF
 $\langle\langle\rho(\e_1)\rho(\e_2)\rangle\rangle=
 \langle\rho(\e_1)\rho(\e_2)\rangle
 -\langle\rho(\e_1)\rangle\langle\rho(\e_2)\rangle$.

%%%%%%%%%%%%%%%%%%%%%%%%%%%%%%%%%%%%%%%%%%%%%%%%%%%%%%%%%%%%%%%%%%%%%%%%%%%%
\subsection{Summary of the result}

 Before entering into the detailed analysis,
 we give an outline of the derivation and the result for reference.
 For perturbation theory, 
 the $Q$-matrix is expanded in powers of the $P$ matrix:
\be
 Q(x) = \Si_z\frac{1+iP}{1-iP}
  = \Si_z\left(1+2iP-2P^2+\cdots\right). \label{Qpert}
\ee
 Correspondingly, the result is expressed by using 
 the expansion of the diffusion propagator \cite{Efetov}
\be
 \Pi(q,\e) = \frac{\Delta}{2\pi}\frac{1}{Dq^2-i\e}. \label{pi}
\ee
 The expansion parameter is $1/g$
 where $g=\pi E_c/\Delta=\pi D/\Delta L^2$
 is the dimensionless conductance.
 It does not appear in the zero-mode sector of the propagator ($q=0$)
 and the expansion is not justified.
 Actually, treating the zero mode exactly (nonperturbatively), 
 and neglecting other nonzero modes, 
 we can obtain the ergodic result.
 The $Q$ matrix for the zero mode is written as 
\be
 Q = T\Si_z\bar{T}, \label{Qerg}
\ee
 where $T$ is independent of the spatial coordinate 
 and its explicit parametrization is given in the following.
 In order to incorporate the zero and nonzero modes into the analysis
 we should use the parametrization
\be
 Q(x) = T\tilde{Q}(x)\bar{T}. \label{Qnonp}
\ee
 $\tilde{Q}$ parametrizes the nonzero modes 
 and is expanded in powers of the $P$-matrix as in Eq.(\ref{Qpert}).
 The zero mode $Q=T\Si_z\bar{T}$ is treated nonperturbatively
 so that the ergodic result is obtained.
 This parametrization is reminiscent of the renormalization group 
 calculation (see, e.g., Ref.\cite{Efetov}) and was used by KM.
 They considered integrations of the nonzero modes first and 
 found corrections to the ergodic result.
 For a technical reason the result was applicable 
 only to the domain $z\ll g$ 
 where $z=\pi\e/\Delta$ is the scaled energy variable.
 Here we consider the zero-mode integration first and then 
 integrate the nonzero modes.
 This method allows us to consider the domain $z\gg 1$ discussed by AA.
 For comparison, we present the KM method in Sec.\ref{KM}.

 The zero-mode model is equivalent to chiral RMT.
 This ergodic limit can be obtained by putting $g=\infty$ 
 in the above functional integral form.
 The result is scaled by the mean level spacing $\Delta$ to give 
\be
 \rho_1(z) &=& \Delta\left<\rho(\e=\Delta z/\pi)\right> 
 = \rho_1^{(0)}(z), \label{doserg} \\ 
 \rho_2(z_1,z_2) &=& 
 \Delta^2\langle\langle\rho(\e_1=\Delta z_1/\pi)\rho(\e_2=\Delta z_2/\pi)
 \rangle\rangle \no\\
 &=& -K^2(z_1,z_2), \label{2pterg}
\ee
 where 
\be
 \rho_1^{(0)}(z) &=& \frac{\pi z}{2}[J_0^2(z)+J_1^2(z)], \no\\
 K(z_1,z_2) &=& \frac{\pi\sqrt{z_1z_2}}{z_1^2-z_2^2}
 [z_1J_1(z_1)J_0(z_2)-z_2J_0(z_1)J_1(z_2)]. \no\\
\ee
 This result does not depend on any parameter and is universal.
 It was obtained in Ref.\cite{VZ} by using the orthogonal polynomial method
 and in Ref.\cite{AST} using the supersymmetry method.

 How is it changed if we include the nonzero modes? 
 If we treat all the modes perturbatively,
 the result is expressed by the diffusion propagator.
 The expansion (\ref{Qpert}) is used to give 
\be
 \left<\rho(\e)\right> 
 &\sim& \frac{1}{\Delta}
 \biggl[1+\frac{1}{2}\mbox{Re}\Bigl(\sum_{q}\Pi(q,\e)\Bigr)^2
 \biggr], \label{dospert} \\
 \langle\langle\rho(\e_1)\rho(\e_2)\rangle\rangle
 &\sim& \frac{1}{2\Delta^2}\mbox{Re}
 \sum_q\bigl[\Pi^2(q,(\e_1+\e_2)/2) \no\\
 & & +\Pi^2(q,(\e_1-\e_2)/2)\bigr]. 
 \label{2ptpert}
\ee
 This expression includes the zero-mode and 
 is justified for $g\gg 1$ and $z\gg 1$.
 The zero mode contribution gives 
 the asymptotic form of the ergodic result 
 as was shown in Ref.\cite{TI}.
 
 Before discussing the exact treatment of the zero mode
 we must mention the effect of 
 the renormalization of the mean level spacing.
 The quantity $\Delta$ was introduced as the mean level spacing 
 at $g=\infty$ and $z=\infty$.
 For traditional symmetry classes, it remains unchanged 
 even if we include the nonzero modes (finite-$g$ effect), 
 which is a consequence of the particle conservation law.
 However, this is not the case in chiral systems. 
 For finite-$g$, the nonzero modes contribute to $\Delta$,
 which means that the DOS is renormalized 
 as was discussed in Ref.\cite{Gade}.
 Referring to Eq.(\ref{dospert}), 
 we define the renormalized mean level spacing  
\be
 \frac{1}{\tilde{\Delta}}
 \sim \frac{1}{\Delta}
 \biggl[1+\frac{1}{2}\mbox{Re}\Bigl(\sum_{q\ne 0}\Pi(q,\e)\Bigr)^2+\cdots
 \biggr]. \label{tildeDelta}
\ee
 Note that the zero mode is excluded in this expression.
 Contributions of the zero mode 
 are totally different from those of other modes.
 The nonzero modes determine 
 the macroscopic behavior of the DOS $1/\tilde{\Delta}$, 
 while the zero mode determines the universal microscopic behavior 
 after scaling in terms of $\tilde{\Delta}$.

 A naive calculation shows that 
 $\tilde{\Delta}$ is divergent in some cases and 
 should be renormalized to a finite value using a regularization. 
 We are interested in the microscopic behavior 
 after the mean level spacing is scaled out. 
 The effect of nonzero modes in the microscopic domain
 is present even after the scaling 
 and we discuss it in the following.

 We turn to the main results in this section.
 We use the parametrization (\ref{Qnonp}) 
 to treat the zero and nonzero modes separately.
 The zero mode is parametrized so that the ergodic results 
 (\ref{doserg}) and (\ref{2pterg}) are reproduced 
 and the nonzero modes are treated perturbatively.
 The domain $z\ll g$ was first considered by KM 
 for nonchiral systems and we call it KM's domain.
 Up to second order in $1/g$, 
 the DOS in the KM's domain is given by 
\be
 \rho_1(z) &=& \tilde{\Delta}
 \langle\rho(\e=\tilde{\Delta}z/\pi)\rangle \no\\
 &\sim&
 \left[1+\frac{a_d}{8g^2}
 \left(2z\frac{d}{dz}+z^2\frac{d^2}{dz^2}\right)
 \right]\rho_1^{(0)}(z). \label{doskm}
\ee
 $a_d$ is the momentum integration
\be
 a_d = \frac{1}{8\pi^4}\sum_{n\ge 0,n^2\ne 0}
 \left(\frac{1}{n^2}\right)^2. \label{a2}
\ee
 We used the periodic boundary condition.
 The TLCF is 
\be
 \rho_2(z_1,z_2) &=& 
 \tilde{\Delta}^2
 \langle\langle\rho(\e_1=\tilde{\Delta}z_1/\pi)
 \rho(\e_2=\tilde{\Delta}z_2/\pi)
 \rangle\rangle \no\\
 &\sim& -\biggl\{\biggl[
 1+\frac{a_d}{8g^2}
 \left(z_1\frac{\partial}{\partial z_1}
 +z_2\frac{\partial}{\partial z_2}\right) \no\\
 & & +\frac{a_d}{8g^2}\left(z_1\frac{\partial}{\partial z_1}
 +z_2\frac{\partial}{\partial z_2}\right)^2
 \biggr]K(z_1,z_2)\biggr\}^2. \no\\\label{2ptkm}
\ee
 The result was scaled by 
 the renormalized mean level spacing (\ref{tildeDelta}).
 The calculation of the DOS for chiral systems has been done 
 in Ref.\cite{TI} 
 but the renormalized mean level spacing was not introduced.
 It leads to a different conclusion on level statistics
 as we discuss in Secs.\ref{KM} and \ref{conc}.

 We now consider the AA domain $z\gg 1$, $g \gg 1$. 
 The scaled DOS is given by
\be
 \rho_1(z) \sim
 1-\frac{\cos 2z}{2z}{\cal D}(z)+\frac{1}{8z^2}, \label{dosaa}
\ee
 where ${\cal D}(z)$ is the spectral determinant
\be
 {\cal D}(z) &=& \prod_{q\ge 0, q^2\ne 0}
 \frac{(Dq^2)^2}{(Dq^2)^2+\e^2} \no\\
 &=& \prod_{n\ge 0, n^2\ne 0}
 \frac{g^2(4\pi^2 n^2)^2}{g^2(4\pi^2 n^2)^2+z^2}. \label{sdet} 
\ee
 The TLCF is
\bw
\be
 \rho_2(z_1,z_2) &\sim& 
 \frac{1}{2}\mbox{Re}\sum_{q^2\ne 0}\left(\Pi_+^2+\Pi_-^2\right) 
 +\frac{\sin 2z_1}{2z_1}{\cal D}_1 \mbox{Im}\sum_{q^2\ne 0}
 \left(\Pi_++\Pi_-\right) 
 +\frac{\sin 2z_2}{2z_2}{\cal D}_2 \mbox{Im}\sum_{q^2\ne 0}
 \left(\Pi_+-\Pi_-\right) \no\\
 & & +\frac{1}{8z_1z_2}\left[
 {\cal D}_1{\cal D}_2({\cal D}_+^{2}{\cal D}_-^{-2}-1)
 \cos 2(z_1+z_2) 
 +{\cal D}_1{\cal D}_2({\cal D}_-^{2}{\cal D}_+^{-2}-1)
 \cos 2(z_1-z_2)
 \right] \no\\
 & & -\frac{1}{2(z_1+z_2)^2}\left[1+
 {\cal D}_1{\cal D}_2{\cal D}_+^2{\cal D}_-^{-2}\cos 2(z_1+z_2)\right]
 -\frac{1}{2(z_1-z_2)^2}\left[1-
 {\cal D}_1{\cal D}_2{\cal D}_-^2{\cal D}_+^{-2}\cos 2(z_1-z_2)\right]  \no\\
 & & +\frac{1}{z_1^2-z_2^2}\left(
 {\cal D}_1  \sin 2z_1-{\cal D}_2 \sin 2z_2\right), 
 \label{2ptaa}
\ee
\ew
 where ${\cal D}_{1,2}={\cal D}(z_{1,2})$, 
 ${\cal D}_{\pm}={\cal D}((z_1\pm z_2)/2)$, and 
 $\Pi_{1,2}=\Pi(q,\e_{1,2})$, $\Pi_\pm=\Pi(q,(\e_1\pm \e_2)/2)$.
 The result is expressed using 
 the spectral determinant as the AA result \cite{AA}.
 Equation (\ref{dosaa}) can be interpreted as follows.
 Consider the asymptotic form of the ergodic result (\ref{doserg})  
\be
 \rho_1(z) \sim 1-\frac{\cos 2z}{2z}+\frac{1}{8z^2}+\cdots. \label{asyerg}
\ee
 Then, including the spectral determinant in the oscillating term, 
 one finds Eq.(\ref{dosaa}). 
 Eq.(\ref{2ptaa}) is more complicated, but 
 we can see that the ergodic limit gives the asymptotic form of 
 the exact result (\ref{2pterg}).
 While standard perturbation theory gives nonoscillating terms,
 expansions around two saddle points \cite{AA}
 are required to get oscillating terms.

 We emphasize that Eqs.(\ref{doskm}), (\ref{2ptkm}), (\ref{dosaa}), 
 and (\ref{2ptaa}) are the main results in this section.
 They have the following properties.

%%%%%%%%%%%%%%%%%%%%%%%%%%%%%%%%%%
 {\it Common domain $1\ll z\ll g$}.
 The KM and AA results have a common domain $1\ll z\ll g$  
 where the asymptotic expansion of the Bessel function and the expansion 
 of the spectral determinant in $z/g$ can be used.
 In this domain, the DOS and TLCF are approximated as 
\be
 \rho_1(z) &\sim& 
 1-\frac{\cos 2z}{2z}+\frac{1}{8z^2}
 +\frac{a_d}{4g^2}z\cos 2z, \\
 \rho_2(z_1,z_2)
 &\sim& 
 -\biggl\{
 \frac{\sin(z_1-z_2)}{z_1-z_2}-\frac{\cos(z_1+z_2)}{z_1+z_2} \no\\
 & & +\frac{a_d}{8g^2}
 (z_1+z_2)\cos (z_1+z_2) \no\\
 & & -\frac{a_d}{8g^2}(z_1-z_2)\sin (z_1-z_2)
 \biggr\}^2.
\ee

%%%%%%%%%%%%%%%%%%%%%%%%%%%%%%%%%%
 {\em Small $z$}.
 At small energies, the expansion of the Bessel function in $z$
 is used in Eqs.(\ref{doskm}) and (\ref{2ptkm}) to give
\be
 \rho_1(z) &\sim&  
 \frac{\pi z}{2} 
 \left(1+\frac{a_d}{4g^2}\right), \\
 \rho_2(z_1,z_2)
 &\sim&  
 -\frac{\pi^2 z_1z_2}{4}\left(1+\frac{a_d}{2g^2}\right). \label{lr}
\ee
 These results show that level repulsion at the origin weakens, 
 which is consistent with the intuitive picture.

%%%%%%%%%%%%%%%%%%%%%%%%%%%%%%%%%%
 {\it Unitary limit}. 
 Taking $z, z_1+z_2\to \infty$, we obtain the unitary limit as 
 $\rho_1(z)\to 1$ and 
\be
 R(z_1,z_2) &=& 1+\frac{\rho_2(z_1,z_2)}
 {\rho_1(z_1)\rho_1(z_2)} \no\\
 &\to& 1+\frac{1}{2}\mbox{Re}\sum_{q^2\ne 0}\Pi_-^2
 -\frac{1}{2(z_1-z_2)^2} \no\\
 & & +\frac{\cos 2(z_1-z_2)}{2(z_1-z_2)^2} 
 {\cal D}\left(\frac{z_1-z_2}{2}\right).
\ee
 This result is consistent with the AA's result \cite{AA} 
 for the unitary class.
 We note the relations
 $\Pi(q,\e/2;g)=2\Pi(q,\e;2g)$ and ${\cal D}(z/2;g)={\cal D}(z;2g)$. 
 The coefficient 2 in front of $g$ originates from chiral symmetry.
 Compared our $\sigma$ model (\ref{F2}) 
 with the model for unitary symmetry \cite{Efetov}, 
 we see the size of the $Q$ matrix is doubled due to chiral symmetry.

%%%%%%%%%%%%%%%%%%%%%%%%%%%%%%%%%%
 {\it $z_1=z_2$}. 
 The relation $\rho_2(z,z) = -\rho_1^2(z)$ holds for arbitrary $g$.
 It can be used to derive the DOS from the TLCF.

%%%%%%%%%%%%%%%%%%%%%%%%%%%%%%%%%%%%%%%%%%%%%%%%%%%%%%%%%%%%%%%%%%%%%%%%%%%%
\subsection{Density of states}
\label{dos}

%%%%%%%%%%%%%%%%%%%%%%%%%%%%%%
\subsubsection{Perturbative calculation}

 Now we go into details of the calculation of the DOS (\ref{rho}).
 The perturbative calculation for nonzero modes
 is considered using the expansion of the $Q$-matrix 
 in $P$ as Eq.(\ref{Qpert}). 
 The $P$ matrix is parametrized for the chiral unitary class as 
\be
 P = \left(\ba{cc} 0 & t \\ t & 0 \ea\right), \quad
 t = \left(\ba{cc} a & \sigma \\ \rho & ib \ea\right),
 \label{P}
\ee
 where $a$, $b$ are real variables, and $\sigma$, $\rho$ Grassmann ones. 
 The measure of this parametrization is normalized to unity. 
 We define the average 
\be
 \left<\cdots \right> &=&
 \int {\cal D}Q \left(\cdots\right)\mbox{e}^{-F_1^{(0)}}, \no\\
 F_1^{(0)} &=& \frac{\pi D}{\Delta V}\int_x \str (\nabla P)^2
 -\frac{i\pi\e}{\Delta V}\int_x \str P^2, \label{F10}
\ee
 where $F_1^{(0)}$ is the second order part of $F_1$.
 Performing expansions in $P$ as
\be
 \left<\rho(\e)\right> &\sim& \frac{1}{\Delta}\mbox{Re}
 \biggl(1-\frac{1}{2V}\int_x\left<\str kP^2\right> \no\\
 & & +\frac{1}{2V}\int_x\left<\str kP^4\right>
 +\cdots\biggr),
\ee
 and using the contraction rules derived 
 in Appendix \ref{cont} as Eq.(\ref{cont1}),
 we obtain the result (\ref{dospert}).

 As we emphasized in the previous subsection, 
 this perturbative calculation of the nonzero modes suggests that 
 the mean level spacing $\tilde{\Delta}$ 
 is renormalized as Eq.(\ref{tildeDelta}).
 The exact definition of $\tilde{\Delta}$ can be written as
\be
 \frac{1}{\tilde{\Delta}} = 
 \frac{1}{\Delta}\int {\cal D}\tilde{Q}
 \left[\frac{1}{4V}\int_x\str k\Si_z\tilde{Q}(x)\right]
 \mbox{e}^{-F_1[\tilde{Q}]}.
\ee 
 Thus effect of the self-interacting diffusion bubble is 
 renormalized to the mean level spacing.
 It corresponds to imposing the constraint 
 $\langle\tilde{Q}\rangle_{F_1}=\Si_z$.
 In Sec.\ref{KM} we give a detailed analysis using the KM method.

%%%%%%%%%%%%%%%%%%%%%%%%%%%%%
\subsubsection{Ergodic limit}

 At the ergodic limit $g\to\infty$, 
 spatial dependence of the $Q$ matrix is neglected 
 and the DOS is reduced to the form
\be
 \rho_1(z) = \frac{1}{4}\int {\cal D}Q\str k\Si_z Q 
 \exp\left(-\frac{iz}{2}\str \Si_z Q\right). \label{rhoerg}
\ee
 Following Ref.\cite{AST}, we parametrize the $Q$ matrix as
\be
 Q &=& T\Si_z\bar{T}, 
 \quad T = UT_0\bar{U}, \no\\
 T_0 &=& 
 \left(\ba{cc} \cos\frac{\hat{\theta}}{2} & -i\sin\frac{\hat{\theta}}{2} \\
 -i\sin\frac{\hat{\theta}}{2} & \cos\frac{\hat{\theta}}{2} \ea\right), \quad
 \hat{\theta} = \left(\ba{cc} \theta_F & 0 \\ 0 & i\theta_B \ea\right), \no\\
 U &=& \left(\ba{cc} u & 0 \\ 0 & u \ea\right), \quad
 u = \exp\left(\ba{cc} 0 & \xi \\ \eta & 0 \ea\right), \label{nonpert}
\ee
 where $-\pi\le\theta_F\le\pi$ and $0\le\theta_B\le\infty$.
 The measure is given by
\be
 {\cal D}Q &=& d\theta_Bd\theta_F d\xi d\eta
 \frac{1}{2\pi} \no\\
 & & \times
 \frac{\cosh\theta_B\cos\theta_F-1-i\sinh\theta_B\sin\theta_F}
 {(\cosh\theta_B-\cos\theta_F)^2}.
\ee
 We note that the compact (noncompact) variable $\theta_F$ ($\theta_B$) 
 is used for the fermion-fermion (boson-boson) block \cite{Efetov}. 
 Substituting this parametrization into Eq.(\ref{rhoerg}) and
 integrating the Grassmann variables, we find 
\be
 \rho_1(z) &=& 
 1+\mbox{Im}
 \int_z^\infty dt\int_0^\infty d\theta_B\int_{0}^\pi d\theta_F 
 \frac{1}{\pi} \no\\
 & & \times (\cosh\theta_B\cos\theta_F-1)
 \mbox{e}^{it^+\left(\cosh\theta_B-\cos\theta_F\right)} \no\\
 &=& 1-\frac{\pi}{2}
 \int_z^\infty dt
 \left(J_0^2(t)-J_1^2(t)\right). \label{rhoergint}
\ee
 Here we introduced the auxiliary variable $t^+=t+i0$
 and assumed $z>0$ ($t>0$).
 For the Bessel function, we used integral representations 
\be
 J_0(z) 
 = \mbox{Re}\frac{-2i}{\pi}\int_0^\infty d\theta_B
 \mbox{e}^{iz^+\cosh\theta_B}, \label{J0nc}
\ee
 for the noncompact variable and 
\be
 J_0(z) = \frac{1}{\pi}\int_0^\pi d\theta_F\mbox{e}^{iz\cos\theta_F},
 \label{J0c}
\ee
 for the compact variable.
 $J_1$ is given by $J_1(z)=-J_0'(z)$.
 Integrating the variable $t$, we obtain Eq.(\ref{doserg}).

 The asymptotic form at $z\gg 1$ is given by Eq.(\ref{asyerg}).
 This cannot be obtained by standard perturbation theory 
 which gives only nonoscillating terms, 
 the first and third terms in Eq.(\ref{asyerg}). 
 The oscillating second term can be obtained by taking into account 
 two saddle points for integrals of $\theta_{B,F}$ in Eq.(\ref{rhoergint}).
 In addition to the ``standard saddle point'' $(\theta_B,\theta_F)=(0,0)$ 
 we have another ``supersymmetry-breaking saddle point'' $(0,\pi)$.
 We note that the point $(0,0)$ corresponds to $Q=\Sigma_z$
 and $(0,\pi)$ to $Q=-k\Sigma_z$.
 This is precisely the idea of the calculation in Ref.\cite{AA}.
 Taking into account fluctuations around these points, 
 we can obtain the desired result.

 In fact this idea is used to find 
 the correct asymptotics of the Bessel function.
 The noncompact representation (\ref{J0nc})
 is used for $\theta_B$ and has the saddle point $\theta_B=0$.
 The compact representation (\ref{J0c}) for $\theta_F$ 
 has the saddle points $\theta_F=0, \pi$.
 Expanding around these saddle points respectively, we have 
\be
 J_0(z) 
 &\sim& \sqrt{\frac{1}{\pi z}}\biggl[
 \left(1-\frac{1}{8z}+\cdots\right)\cos z \no\\
 & & +\left(1+\frac{1}{8z}+\cdots\right)\sin z\biggr].  \label{J0asy}
\ee
 It is interesting to note that 
 the expansion around a single saddle point 
 is required for the noncompact representation (\ref{J0nc})
 and two points for the compact representation (\ref{J0c}).
 When Eq.(\ref{J0nc}) is deformed to Eq.(\ref{J0c}) 
 the single point $\theta_B=0$ splits into 
 the two points $\theta_F=0, \pi$.
 It can be shown by considering the deformation of the integral contour 
 used in Ref.\cite{AST}.
 We find
\be
 \frac{-2i}{\pi}\int_0^{\infty}d\theta\mbox{e}^{iz^+\cosh\theta} 
 &=& 
 \frac{2}{\pi}\int_0^{\pi/2}d\theta\mbox{e}^{iz\cos\theta} \no\\
 & & -\frac{2i}{\pi}\int_0^{\infty}d\theta
 \mbox{e}^{-z\sinh\theta}. \label{H0}
\ee
 This representation is known as the Hankel function $H_0=J_0+iN_0$.
 Taking the real part, we obtain 
\be
 J_0(z) = \frac{1}{\pi}\int_0^{\pi/2}d\theta\mbox{e}^{iz\cos\theta}
 +\frac{1}{\pi}\int_0^{\pi/2}d\theta\mbox{e}^{-iz\cos\theta}.
\ee
 This expression is reduced to Eq.(\ref{J0c}) by changing the variable 
 $\theta\to \pi-\theta$ in the second term.
 Thus the point 0 in the second term is changed to $\pi$.
 Note that the real part of the integral is taken 
 in the noncompact representation (\ref{J0nc}),
 which gives the second term.

 This method, taking into account a set of nontrivial saddle points, 
 is the main idea of the nonperturbative calculation.
 It produces the exact result for the unitary class 
 and the asymptotic ones for the orthogonal and symplectic classes \cite{AA}.
 It has been used even for 
 the replica \cite{replica} and Keldysh \cite{keld} $\sigma$ models.
 For chiral symmetric systems at the ergodic limit, 
 a similar technique has been used in Ref.\cite{NGK}
 to find the asymptotic result (\ref{asyerg}).
 In the following, 
 we examine how the effect of nonzero modes 
 is incorporated into the asymptotic form.

%%%%%%%%%%%%%%%%%%%%%%%%%%%%%
\subsubsection{Integration of the zero mode}

 We write the $Q$ matrix as Eq.(\ref{Qnonp}) 
 and use the parametrizations (\ref{Qpert}) and (\ref{P}) for $\tilde{Q}$,
 and (\ref{nonpert}) for $T$.
 It is slightly modified as 
\be
 Q(x) = UT_0\bar{U}\tilde{Q}(x)U\bar{T}_0\bar{U}
 \to UT_0\tilde{Q}(x)\bar{T}_0\bar{U}.
\ee
 As a result, $F_1$ becomes independent of
 the Grassmann variables of the zero mode.
 The preexponential term in Eq.(\ref{rho}) is written 
 explicitly using the Grassmann variables as
\be
 \str k\Si_z Q(x) &\to& \str k\Si_z T_0\tilde{Q}(x)\bar{T}_0 \no\\
 & & +2\xi\eta\str\Si_z T_0\tilde{Q}(x)\bar{T}_0. 
\ee
 We neglected contributions that 
 vanish after integrations over $\xi$ and $\eta$.
 The first term does not include those variables and 
 we can put $T_0=1$ for the integrations. 
 The second term is also easily integrated out and 
 we thus have 
 $\left<\rho(\e)\right> =
 \left<\rho(\e)\right>_1+\left<\rho(\e)\right>_2$
 where 
\bw
\be
 \left<\rho(\e)\right>_1 &=& 
 \frac{1}{4\Delta V}\mbox{Re}\int {\cal D}\tilde{Q}{\cal J}[\tilde{Q}]
 \left[\int\str k\Si_z\tilde{Q}\right] \mbox{e}^{-F_1[\tilde{Q}]}, \no\\
 \left<\rho(\e)\right>_2 &=& 
 \frac{1}{2\pi\Delta}\mbox{Re}
 \int_0^\infty d\theta_B \int_{-\pi}^\pi d\theta_F 
 \frac{(\cosh\theta_B\cos\theta_F-1+i\sinh\theta_B\sin\theta_F)}
 {(\cosh\theta_B-\cos\theta_F)^2}I(\e,\theta_B,\theta_F), \no\\
 I(\e,\theta_B,\theta_F) 
 &=& -i\frac{\partial}{\partial z}\int {\cal D}\tilde{Q}{\cal J}[\tilde{Q}]
 \exp\left[
 -\frac{\pi D}{4\Delta V}\int \str (\nabla \tilde{Q})^2 
 -\frac{i\pi\e}{2\Delta V}\int \str \Si_z T_0\tilde{Q}\bar{T}_0\right].
\ee
\ew
 $\left<\rho(\e)\right>_1$ gives the perturbative result 
 (\ref{dospert}) without the zero mode contribution 
 and is equal to the inverse of 
 the renormalized mean level spacing $1/\tilde{\Delta}$.
 $\left<\rho(\e)\right>_2$ includes the ergodic result 
 and is nonperturbative.

 We note that the Jacobian ${\cal J}[\tilde{Q}]$ contribution 
 exists in the present parametrization (\ref{Qnonp}).
 It depends on the nonzero modes only and can be written as
\be
 {\cal J}[\tilde{Q}] = \exp\left[\frac{1}{4V}\int
 (\str P\Si_x)^2 +O(P^4) \right]. \label{jacobian}
\ee
 This contribution changes the renormalized mean level spacing slightly 
 and the scaled DOS $\rho_1(z)$ is not changed in our approximation.
 For this reason, 
 we neglect this contribution in the present section.
 It is treated in Sec.\ref{double} 
 when we discuss the DOS renormalization.
 
 Let us turn to the calculation of $\left<\rho(\e)\right>_2$. 
 The kinetic term in $F_1$ does not include the zero mode
 and is expanded in powers of $P$.
 The second term in $F_1$ (and the preexponential term) is expanded as
\be
 & & -\frac{1}{2}\int \str \Si_z T_0\tilde{Q}\bar{T}_0 \no\\
 &=& V(\cosh\theta_B-\cos\theta_F) \no\\ 
 & & +\int \left[
 \str \left(k_FP^2\right)\cos\theta_F
 +\str \left(k_BP^2\right)\cosh\theta_B\right] \no\\
 & & -\int \bigl[
 \str \left(k_F\Si_x P^3\right)\sin\theta_F 
 -i\str \left(k_B\Si_x P^3\right)\sinh\theta_B\bigr] \no\\
 & & +\cdots,
\ee
 where $k_{F,B}=(1\pm k)/2$.
 In the following calculation 
 we neglect odd terms in the $P$-matrix.
 Their contributions give $1/g^3$-corrections at most.
 Another reason to neglect them is that
 they involve a factor $\sin\hat{\theta}$ 
 which goes to zero 
 at the saddle points $\theta_{F}=0, \pi$ and $\theta_B=0$.
 Using this approximation, we find the simplified expression
\be
 I(\e,\theta_B,\theta_F)
 &\sim& -i\frac{\partial}{\partial z}\mbox{e}^{iz(\lambda_B-\lambda_F)}
 \left<\mbox{e}^{iz\lambda_B A_B+iz\lambda_F A_F}
 \right>_{\rm kin}, \no\\
 A_{F,B}[\tilde{Q}] &=& 
 -\frac{1}{2V}\int \str k_{F,B}\Si_z [\tilde{Q}-\Si_z], \no\\
 \left<\cdots\right>_{\rm kin} &=& \int {\cal D}\tilde{Q}
 \left(\cdots\right)\mbox{e}^{-F_{\rm kin}}, \no\\
 F_{\rm kin} &=&
 \frac{\pi D}{4\Delta V}\int \str (\nabla \tilde{Q})^2, 
\ee
 where $z=\pi\e/\Delta$, 
 $\lambda_B=\cosh\theta_B$, and $\lambda_F=\cos\theta_F$.
 $A_{F,B}[\tilde{Q}]$ include even powers in $P$.
 Introducing the auxiliary variable $t$, we obtain
\be
 & & \left<\rho(\e)\right>_2  = 
 \frac{1}{\pi\Delta}\mbox{Im}
 \int_z^\infty dt\int_0^\infty d\theta_B \int_{0}^\pi d\theta_F 
 (\lambda_B\lambda_F-1)  \no\\
 & & \times
 \mbox{e}^{it^+(\lambda_B-\lambda_F)}
 \left[1-(t-z)\frac{\partial}{\partial z}\right] 
 \left<\mbox{e}^{iz\lambda_B A_B+iz\lambda_F A_F}\right>_{\rm kin}. \no\\
\ee

 Now the problem is how integrations of 
 the variables $\theta_{B,F}$ are performed.
 They can be done by noting that 
 the variable $t$ in the exponential is shifted 
 to $t^++zA_B$ or $t-zA_F$ compared with the ergodic limit.
 For the fermion part $\theta_F$, 
 there is no convergence problem
 and the Bessel function is derived.
 It is also the case for the boson part $\theta_B$  
 since the convergence problem does not arise 
 for the part including Grassmann variables 
 and the other parts are real.
 The only difference is that we cannot take the real part 
 for the expression after integration of $\theta_B$
 since the argument $t+zA_B$ includes Grassmann variables.
 We get the Hankel function $H_0=J_0+iN_0$ 
 instead of the Bessel function $J_0$ [see Eq.(\ref{H0})].
 However, the imaginary part $iN_0$ does not 
 contribute to the final result since the functional $A_B$
 is reduced to a real function in the end.
 This is valid in our approximation keeping 
 contributions up to second order in $1/g$.
 Thus we neglect the imaginary part and obtain 
\be
 \left<\rho(\e)\right>_2 &\sim& 
 \frac{\pi}{2\Delta}\mbox{Re}\frac{d}{dz}
 \int_z^\infty dt(t-z) \no\\
 & & \times \Bigl<J_0(t+zA_B)J_0(t-zA_F) \no\\
 & & -J_1(t+zA_B)J_1(t-zA_F)\Bigr>_{\rm kin}.
 \label{rho2}
\ee
 The ergodic limit $g=\infty$ can be found easily 
 by putting $A_{F,B}[\tilde{Q}]=0$.
 We note again that this equation was obtained 
 by neglecting contributions including 
 $\sin\theta_F$ or $\sinh\theta_B$.
 This approximation is valid up to second order in $1/g$.
 It still remains to carry out integrations over the nonzero modes. 
 In the following we consider two limiting cases.

%%%%%%%%%%%%%%%
\subsubsection{KM's domain ($z\ll g$)} 

 The case $z\ll g$ can be considered using KM's method \cite{KM}. 
 For chiral systems, it was considered in Ref.\cite{TI}.
 In our method, the Bessel functions in Eq.(\ref{rho2}) 
 are expanded in powers of $zA_{F,B}\sim O(z/g)$ to find
\be
 \left<\rho(\e)\right>_2 &\sim& 
 \frac{1}{\Delta}\mbox{Re}
 \biggl[1+\frac{1}{2}\left<A_B-A_F\right>_{\rm kin}\frac{d}{dz}z \no\\
 & & +\frac{1}{8}\left<(A_B-A_F)^2\right>_{\rm kin}
 \frac{d}{dz}z^2\frac{d}{dz}\biggr]
 \left[\rho_1^{(0)}(z)-1\right]. \no\\
\ee
 Combining with the perturbative contribution
\be
 \left<\rho(\e)\right>_1 &\sim& 
 \frac{1}{\Delta}\mbox{Re}
 \biggl[1+\frac{1}{2}\left<A_B-A_F\right>_{\rm kin}+\cdots\biggr],
\ee
 we find
\be
 \left<\rho(\e)\right> &\sim& 
 \frac{1}{\Delta}\mbox{Re}
 \biggl[1+\frac{1}{2}\left<A_B-A_F\right>_{\rm kin}\frac{d}{dz}z \no\\
 & & +\frac{1}{8}\left<(A_B-A_F)^2\right>_{\rm kin}
 \frac{d}{dz}z^2\frac{d}{dz}\biggr]
 \rho_1^{(0)}(z). \label{barerho}
\ee
 Up to here the DOS is scaled 
 in terms of the bare mean level spacing $\Delta$.
 We introduce the renormalized mean level spacing as 
 $1/\tilde{\Delta} = \left<\rho(\e)\right>_1$.
 Defining the energy variable as $\tilde{z}=\pi\e/\tilde{\Delta}$,
 we use the transformation formula for a function $f(z)$
\be
 f(z) &=& \left[1+\Bigl(\frac{\tilde{\Delta}}{\Delta}-1\Bigr)
 \tilde{z}\frac{d}{d\tilde{z}}+\cdots\right]
 f(\tilde{z}) \no\\
 &\sim& \left[1-\frac{1}{2}\left<A_B-A_F\right>_{\rm kin}
 \tilde{z}\frac{d}{d\tilde{z}}+\cdots\right]
 f(\tilde{z}). \label{tr}
\ee
 It is applied to Eq.(\ref{barerho}) to find 
\be
 \rho_1(\tilde{z}) &=& \tilde{\Delta}
 \langle\rho(\e=\tilde{\Delta}\tilde{z}/\pi)\rangle \no\\
 &\sim&
 \mbox{Re}
 \biggl[1+\frac{1}{8}\langle\langle(A_B-A_F)^2\rangle\rangle_{\rm kin}
 \frac{d}{d\tilde{z}}\tilde{z}^2\frac{d}{d\tilde{z}}\biggr]
 \rho_1^{(0)}(\tilde{z}), \no\\
\ee
 where
\be
 \langle\langle(A_B-A_F)^2\rangle\rangle_{\rm kin} &=& 
 \left<(A_B-A_F)^2\right>_{\rm kin}-\left<A_B-A_F\right>_{\rm kin}^2 \no\\
 &\sim& \frac{a_d}{g^2}.
\ee
 This is obtained by expanding $\tilde{Q}$ in powers of $P$ 
 and using the contraction (\ref{cont1}) with $\e=0$.
 $a_d$ is momentum summation and is given by Eq.(\ref{a2}).
 Thus we obtain Eq.(\ref{doskm}).

%%%%%%%%%%%%%%%
\subsubsection{AA's domain ($1\ll z$)} 

 In the limit $1\ll z$,
 the asymptotic form of the Bessel function (\ref{J0asy}) 
 is used to write 
\be
 \left<\rho(\e)\right>_2 &\sim& 
 \frac{1}{\Delta}\mbox{Re}\frac{d}{dz}
 \int_z^\infty dt  
 \biggl[
 -\frac{t-z}{4t^3}\left<\mbox{e}^{iz(A_B+A_F)}
 \right>_{\rm kin} \no\\
 & & -i\frac{t-z}{t}
 \left<\mbox{e}^{2it+iz(A_B-A_F)}\right>_{\rm kin}\biggr] \no\\
 &\sim& 
 \frac{1}{\Delta}\mbox{Re}\left[
 \frac{1}{8z^2}{\cal D}(z,1,1)
 -\frac{1}{2z}
 \mbox{e}^{2iz}{\cal D}(z,1,-1)\right], \no\\
 \label{rho2aa}
\ee
 where
\be
 {\cal D}(z,\lambda_B,\lambda_F) &=& \int{\cal D}\tilde{Q}
 \mbox{e}^{-F(z,\lambda_B,\lambda_F)}, \no\\
 F(z,\lambda_B,\lambda_F) &=& F_{\rm kin} 
 +\frac{iz\lambda_F}{2V}\int\str k_F\Si_z (\tilde{Q}-\Si_z) \no\\
 & & +\frac{iz\lambda_B}{2V}\int\str k_B\Si_z (\tilde{Q}-\Si_z).
\ee
 $F(z,1,1)=F_1$ does not break supersymmetry, 
 which means it does not 
 include the supermatrix $k=\mbox{diag}(1,-1)$.
 As a result we obtain ${\cal D}(z,1,1)=1$.
 On the other hand, 
 $F(z,1,-1)$ breaks supersymmetry and the function
 ${\cal D}(z,1,-1)$ is not normalized to unity.
 It is calculated as ${\cal D}(z,1,-1)\sim {\cal D}(z)$,
 where the spectral determinant ${\cal D}(z)$ is given by Eq.(\ref{sdet}).
 We used the approximation of keeping second order in $P$ for $F(z,1,-1)$.
 We refer to Appendix \ref{cont} for details 
 (see also the following paragraph).

 Eq.(\ref{rho2aa}) is rewritten in terms of the energy variable 
 scaled by the renormalized mean level spacing 
 $\tilde{z}=\pi\e/\tilde{\Delta}$.
 We use the formula (\ref{tr}) and  
 the difference between $\Delta$ and $\tilde{\Delta}$ is expressed 
 by the diffusion propagator $\Pi(q,\e)$.
 It represents the self-interacting diffusion bubble 
 and should be canceled out.
 Actually we have contributions 
 from the function ${\cal D}(z,1,-1)$
 by keeping higher-order terms in $P$.
 We find 
\be
 {\cal D}(z,1,-1) &\sim& \int{\cal D}\tilde{Q}
 \mbox{e}^{-F^{(2)}(z,1,-1)} \left(1-F^{(4)}(z,1,-1)\right) \no\\
 &=& {\cal D}(z)\biggl[1+iz\mbox{Re}
 \Bigl(\sum_{q\ne 0}\Pi(q,\e)\Bigr)^2
 \biggr],
\ee
 where $F^{(n)}$ denotes $n$th order part in the expansion.
 The second term cancels with a contribution coming 
 from the transformation (\ref{tr}).
 Noting ${\cal D}(z,g)={\cal D}(\tilde{z},\tilde{g})$, 
 where $g=\pi E_c/\Delta$ and $\tilde{g}=\pi E_c/\tilde{\Delta}$, 
 we finally obtain the result Eq.(\ref{dosaa}) for $1\ll z$.

%%%%%%%%%%%%%%%%%%%%%%%%%%%%%%%%%%%%%%%%%%%%%%%%%%%%%%%%%%%%%%%%%%%%%%%%%%%%
\subsection{Comparison with the KM's method}
\label{KM}

 The obtained result (\ref{doskm}) for the KM's domain 
 differs slightly from Eq.(21) in Ref.\cite{TI}
 by the presence of momentum integration of the propagator 
 $\sum_{q\ne 0}\Pi(q,0)$.
 As we can understand from Eq.(\ref{tildeDelta}),  
 the difference comes from 
 the introduction of the renormalized mean level spacing 
 [Eq.(\ref{barerho}) coincides with Eq.(21) in Ref.\cite{TI}].
 It is expressed as 
 a self-interacting diffusion diagrams 
 [it can be understood by noting the coordinate representation 
 $\sum_q\Pi(q)=\Pi(x,x)$]
 and is renormalized to the mean level spacing.
 In order to make this difference clear,
 we repeat the calculation 
 using the KM's method considered in Ref.\cite{TI}.
 In this method, the nonzero modes are integrated out 
 while keeping the zero mode variables.
 It allows us to obtain the renormalized effective zero-mode action
 and is useful to understand how 
 we can introduce the renormalized mean level spacing.

 We start from the functional for the DOS with the source term
\be
 F &=& \frac{\pi D}{4\Delta V}\int\str [\nabla Q(x)]^2
 +\frac{i\pi\e}{2\Delta V}\int\str Q(x)\Si_z \no\\
 & & +\frac{i\pi J}{2\Delta V}\int\str k Q(x)\Si_z.
\ee
 The $Q$-matrix parametrization (\ref{Qnonp}) is substituted and 
 the nonzero modes $\tilde{Q}$ are expanded in $P$ as Eq.(\ref{Qpert}).
 In our approximation keeping second order in $1/g$, 
 the expansion is performed up to fourth order in $P$.
 The functional $F$ consists of four parts: 
\be
 F &=& F_0+\tilde{F}+F_I+F_J.
\ee
 $F_0$ is the zero mode part $F_0=F[Q=T\Sigma_z\bar{T}]$, 
 $\tilde{F}$ the nonzero mode part $\tilde{F}=F[\tilde{Q}]$, 
 $F_I$ the mixing part, and 
 $F_J$ the source term.
 They are expanded in $P$ as 
\be
 \tilde{F} &=& \tilde{F}^{(2)}+\tilde{F}^{(4)}+\cdots, \no\\
 F_I &=& F_I^{(2)}+F_I^{(3)}+F_I^{(4)}+\cdots, \no\\ 
 F_J &=& F_J^{(0)}+F_J^{(2)}+F_J^{(3)}+F_J^{(4)}+\cdots,  
\ee
 where $F^{(n)}$ denotes the $n$th order part in $P$.

 The effective functional is obtained by integrating the nonzero modes.
 We define $F_{\rm eff}$ as 
\be
 \mbox{e}^{-F_{\rm eff}} &=& \int{\cal D}\tilde{Q}\mbox{e}^{-F} \no\\
 &=& \mbox{e}^{-F_0}\left<
 \mbox{e}^{-\tilde{F}^{(4)}+\cdots-F_I-F_J}\right>_{\tilde{F}^{(2)}},
\ee
 where the average is performed with respect to $\tilde{F}^{(2)}$.
 We use the contraction rules derived in Appendix \ref{cont}.
 Up to second order in cumulant expansion, 
\bw
\be
 F_{\rm eff} &\sim& F_0+F_J^{(0)}
 +\langle F_I^{(4)}\rangle+\langle F_J^{(4)}\rangle 
 -\frac{1}{2}\langle\langle F_I^{(2)2}\rangle\rangle
 -\langle\langle F_I^{(2)}F_J^{(2)}\rangle\rangle \no\\
 &=& \frac{i\pi\e}{2\Delta}
 \biggl[1+\frac{1}{2}\Bigl(\sum_{q\ne 0}\Pi(q,\e)\Bigr)^2\biggr]
 \str Q\Sigma_z 
 +\frac{\pi^2\e^2}{8\Delta^2}\Bigl(\sum_{q\ne 0}\Pi^2(q,\e)\Bigr)
 \left(\str Q\Sigma_z\right)^2 \no\\
 & & +\frac{i\pi J}{2\Delta}
 \biggl[1+\frac{1}{2}\Bigl(\sum_{q\ne 0}\Pi(q,\e)\Bigr)^2\biggr]
 \str k\Sigma_z Q 
  +\frac{\pi^2\e J}{4\Delta^2}\Bigl(\sum_{q\ne 0}\Pi^2(q,\e)\Bigr)
 \str Q\Sigma_z \str k\Sigma_z Q.
\ee
\ew
 Since momentum summations potentially 
 involve divergence, this expansion is somewhat cumbersome.
 This can be clearly seen by 
 considering the KM domain $z\ll g$.
 Then the energy $\e$ in the propagator is neglected 
 in our approximation $\Pi(q,\e)\sim\Pi(q,0)$ and 
 the effective functional can be written as 
\be
 F_{\rm eff} &\sim& \frac{i\pi\e}{2\Delta}
 \biggl[1+\frac{a_d^{(1)2}}{8g^2}\biggr]
 \str Q\Sigma_z 
 +\frac{\pi^2\e^2}{8\Delta^2}\frac{a_d}{4g^2}
 \left(\str Q\Sigma_z\right)^2 \no\\
 & & +\frac{i\pi J}{2\Delta}
 \biggl[1+\frac{a_d^{(1)2}}{8g^2}\biggr]
 \str k\Sigma_z Q \no\\
 & & +\frac{\pi^2\e J}{4\Delta^2}\frac{a_d}{4g^2}
 \str Q\Sigma_z \str k\Sigma_z Q, \label{Feff}
\ee
 where $a_d$ is given by Eq.(\ref{a2}) and 
\be
 a_d^{(1)} &=& \frac{1}{\pi^2}\sum_{n_i\ge 0,n^2\ne 0}\frac{1}{n^2}.
\ee
 This summation is divergent at $d\ge 2$ and we need some regularization.
 Fortunately, and as it should be, 
 the quantity $a_d^{(1)}$ can be renormalized to the mean 
 level spacing by defining the renormalized spacing
\be
 \frac{1}{\tilde{\Delta}} &=& \frac{1}{\Delta}\biggl[
 1+\frac{a_d^{(1)2}}{8g^2}+O(1/g^3)\biggr]. \label{tildeDeltaKM}
\ee
 This is nothing but the expression (\ref{tildeDelta}) at the KM domain.
 $a_d^{(1)}$ corrections come from the average $\langle Q(x)\rangle$. 
 On the other hand the second and fourth term in Eq.(\ref{Feff})
 come from the contraction 
 $\langle\langle Q(x)Q(y)\rangle\rangle$
 and cannot be renormalized to $\Delta$.
 They give the corrections obtained in Eq.(\ref{doskm}).

 Thus the KM's method makes 
 the problem of the renormalization transparent.
 The idea of integrating out fast variables 
 matches the philosophy of the renormalization.
 Nevertheless, we did not use this method for the reason 
 that it is not convenient for calculations in the AA domain $z\gg 1$. 
 Integrations of zero-mode variables 
 naturally bring contributions from nontrivial saddle points, 
 which is an important idea for nonperturbative calculations. 

%%%%%%%%%%%%%%%%%%%%%%%%%%%%%%%%%%%%%%%%%%%%%%%%%%%%%%%%%%%%%%%%%%%%%%%%%%%%
\subsection{Two-level correlation function}
\label{2pt}

 Now we turn to the calculation of the TLCF (\ref{TLCF}).
 $Q$ is an 8$\times$8 supermatrix and 
 the explicit parametrization is different from the previous case.

 For the standard perturbative calculation, 
 we use the expansion Eq.(\ref{Qpert}).
 The explicit parametrization of the $P$ matrix is given by 
\be
 & & P = \left(\ba{cc} 0 & t \\ t & 0 \ea\right), \quad
 t = \left(\ba{cc} t_1 & t_{12} \\ t_{21} & t_2 \ea\right), 
 \no \\
 & & t_1 = \left(\ba{cc} a_1 & \sigma_1 \\ 
 \rho_1 & ib_1 \ea\right), \quad
 t_2 = \left(\ba{cc} a_2 & \sigma_2 \\ 
 \rho_2 & ib_2 \ea\right), \no\\
 & & t_{12} = \left(\ba{cc} c & i\eta \\ 
 \xi^* & id \ea\right), \quad
 t_{21} = \left(\ba{cc} c^* & \xi \\ 
 i\eta^* & id^* \ea\right). \label{P2}
\ee
 $a_{1,2}$, $b_{1,2}$ are real variables, 
 $c$, $d$ complex variables, and 
 the greek symbols denote Grassmann variables.
 As the explicit parametrization implies, 
 $t_{1,2}$ represent the ``chiral'' part and 
 $t_{12,21}$ the ``unitary'' part.
 Starting from the expression (\ref{W}), we have 
\be
 W(\e_1,\e_2) &=& \frac{1}{\Delta^2}\int{\cal D}Q
 \mbox{e}^{-F_2^{(0)}(z_1,z_1,z_2,z_2)+\cdots} \no\\
 & & \times \biggl[1-\frac{1}{2V}\int_x\str k\Lambda_1P^2(x)
 +\cdots\biggr] \no\\
 & & \times \biggl[1-\frac{1}{2V}\int_y\str k\Lambda_2P^2(y)+\cdots\biggr].
\ee
 $F_2^{(2)}(z_1,z_1,z_2,z_2)$ given by Eq.(\ref{F20k}) 
 is second order in $P$ 
 and is the base of the perturbative expansion.
 The contraction rule given by Eq.(\ref{cont2}) 
 is used to evaluate the above expression.
 The leading order contribution to the connected part 
 comes from the contraction
\be
 \left<\str k\Lambda_1P^2(x)\str k\Lambda_2P^2(y)\right>
 = 4\Pi^2(x-y,(\e_1+\e_2)/2). \no\\
\ee
 Thus we obtain the result (\ref{2ptpert})
 which is valid at $g\gg 1$ and $z_{1,2}=\pi\e_{1,2}/\Delta\gg 1$.

 The ergodic limit $g\to\infty$ was considered in Ref.\cite{AST}.
 The $Q$ matrix is parametrized as
\be
 Q = T\Si_z \bar{T}, \quad T = T_{\rm ch}T_{\rm u}.
\ee
 $T_{\rm ch}$ is the chiral part
\bw
\be
 T_{\rm ch} &=& U_{\rm ch}T_{\rm ch}^{(0)}\bar{U}_{\rm ch}, \no\\
 T_{\rm ch}^{(0)} 
 &=& \left(\ba{cccc} 
 \cos\frac{\hat{\theta}_1}{2} & 0 & -i\sin\frac{\hat{\theta}_1}{2} & 0 \\
 0 & \cos\frac{\hat{\theta}_2}{2} & 0 & -i\sin\frac{\hat{\theta}_2}{2} \\
 -i\sin\frac{\hat{\theta}_1}{2} & 0 & \cos\frac{\hat{\theta}_1}{2} & 0 \\
 0 & -i\sin\frac{\hat{\theta}_2}{2} & 
 0 & \cos\frac{\hat{\theta}_2}{2} \ea\right), \quad
 \hat{\theta} =
 \left(\ba{cc} \hat{\theta}_1 & 0 \\ 0 & \hat{\theta}_2 \ea\right) 
 = \left(\ba{cccc} 
 \theta_{1F} & 0 & 0 & 0 \\ 
 0 & i\theta_{1B} & 0 & 0 \\
 0 & 0 & \theta_{2F} & 0 \\ 
 0 & 0 & 0 & i\theta_{2B} \ea\right), \no\\
 U_{\rm ch} &=& \left(\ba{cccc} 
 u_{\rm ch1} & 0 & 0 & 0 \\ 
 0 & u_{\rm ch2} & 0 & 0 \\
 0 & 0 & u_{\rm ch1} & 0 \\
 0 & 0 & 0 & u_{\rm ch2} \\
 \ea\right), \quad
 u_{\rm ch1,2} = \exp\left(\ba{cc} 
 0 & \sigma_{1,2}  \\ 
 \rho_{1,2} & 0 \ea\right),
\ee
 and $T_{\rm u}$ the unitary part
\be
 T_{\rm u} &=& U_{\rm u}T_{\rm u}^{(0)}\bar{U}_{\rm u}, \no\\
 T_{\rm u}^{(0)} &=& \left(\ba{cccc} 
 \cos\frac{\hat{\Omega}}{2} & 0 & 0 & 
 -i\mbox{e}^{i\hat{\varphi}}\sin\frac{\hat{\Omega}}{2} \\
 0 & \cos\frac{\hat{\Omega}}{2} & 
 -i\mbox{e}^{-i\hat{\varphi}}\sin\frac{\hat{\Omega}}{2} & 0 \\
 0 & -i\mbox{e}^{i\hat{\varphi}}\sin\frac{\hat{\Omega}}{2} & 
 \cos\frac{\hat{\Omega}}{2} & 0 \\
 -i\mbox{e}^{-i\hat{\varphi}}\sin\frac{\hat{\Omega}}{2} & 0 & 0 & 
 \cos\frac{\hat{\Omega}}{2} 
 \ea\right), \quad
 \hat{\Omega} = \left(\ba{cc} \Omega_F & 0 \\ 0 & i\Omega_B \ea\right), \quad
 \hat{\varphi} \ =\ 
 \left(\ba{cc} \varphi_F & 0 \\ 0 & \varphi_B \ea\right), \no\\
 U_{\rm u} &=& \left(\ba{cccc} 
 u_{\rm u1} & 0 & 0 & 0 \\ 
 0 & u_{\rm u2} & 0 & 0 \\ 
 0 & 0 & u_{\rm u1} & 0 \\ 
 0 & 0 & 0 & u_{\rm u2} \ea\right), \quad
 u_{\rm u1} = \exp\left(\ba{cc} 0 & \xi \\ -\xi^* & 0 \ea\right), \quad
 u_{\rm u2} =  \exp\left(\ba{cc} 0 & i\eta \\ -i\eta^* & 0 \ea\right).
\ee
 $\sigma$, $\rho$, $\xi$, and $\eta$ are Grassmann variables.
 The integration ranges of the real variables 
 $\theta$, $\Omega$, and $\varphi$ are chosen properly 
 according to the compact or noncompact parametrization \cite{AST}.
 The measure is given by
\be
 {\cal D}Q &=& 
 d\theta_{1B}d\theta_{1F}d\sigma_1d\rho_1
 \frac{1}{4\pi}\frac{\cosh\theta_{1B}\cos\theta_{1F}-1}
 {(\cosh\theta_{1B}-\cos\theta_{1F})^2}
 d\theta_{2B}d\theta_{2F}d\sigma_2d\rho_2
 \frac{1}{4\pi}\frac{\cosh\theta_{2B}\cos\theta_{2F}-1}
 {(\cosh\theta_{2B}-\cos\theta_{2F})^2} \no\\
 & & \times d\Omega_Bd\Omega_F\frac{d\varphi_B}{2\pi}\frac{d\varphi_F}{2\pi} 
 d\xi d\xi^* d\eta^*d\eta
 \frac{\sinh\Omega_B\sin\Omega_F}{(\cosh\Omega_B-\cos\Omega_F)^2}
 \frac{4\cosh\Omega_B\cos\Omega_F}{(\cosh\Omega_B+\cos\Omega_F)^2}.
\ee
 Using this parametrization, 
 after a laborious calculation, 
 we can obtain Eq.(\ref{2pterg}) (see Ref.\cite{AST} for the details).

 The nonperturbative calculation 
 using the parametrization (\ref{Qnonp}) 
 can be done in the same way as that of the DOS. 
 First we integrate the zero-mode variables.
 The details are presented in Appendix \ref{calc2}, 
 and we find for $W$
\be
 W &=& W_1+W_2, 
 \no\\
 W_1 &\sim& \frac{1}{\Delta^2}\biggl<
 \biggl\{\left[1+\frac{1}{2}(A_{B1}-A_{F1})\right]
 \mbox{e}^{iz_1(A_{-1}+A_{+1})} \no\\
 & &
 +\frac{\pi}{2}\frac{\partial}{\partial z_1}\int_{z_1}^\infty dt_1 (t_1-z_1) 
 \left[J_0(t_1+z_1A_{B1})J_0(t_1-z_1A_{F1})
 -J_1(t_1+z_1A_{B1})J_1(t_1-z_1A_{F1})\right] \biggr\} \no\\
 & & \times
 \biggl\{\left[1+\frac{1}{2}(A_{B2}-A_{F2})\right]
 \mbox{e}^{iz_2(A_{B2}+A_{F2})} \no\\
 & & 
 +\frac{\pi}{2}\frac{\partial}{\partial z_2}\int_{z_2}^\infty dt_2 (t_2-z_2)
 \left[J_0(t_2+z_2A_{B2})J_0(t_2-z_2A_{F2})
 -J_1(t_2+z_2A_{B2})J_1(t_2-z_2A_{F2})\right] \biggr\}
 \biggr>_{\rm kin}, 
 \no\\
 W_2 &\sim& \frac{1}{\Delta^2}\int ds_1ds_2
 \frac{4s_1s_2}{(s_1^2-s_2^2)^2}I(z,s), 
 \no\\
 I(z,s) &=& \frac{\pi^2}{4}\biggl<
 z_1z_2(s_1-s_2+C_1)(s_1+s_2-D_1)
 J_0\left(z_1s_1+z_1\frac{C_1-D_1}{2}\right)
 J_0\left(z_1s_2-z_1\frac{C_1+D_1}{2}\right) \no\\
 & & \times 
 (s_1-s_2+C_2)(s_1+s_2-D_2)
 J_0\left(z_2s_1+z_2\frac{C_2-D_2}{2}\right)
 J_0\left(z_2s_2-z_2\frac{C_2+D_2}{2}\right)
 \biggr>_{\rm kin}, \label{W12}
\ee
\ew
 where $s_1=\cosh\Omega_B$, $s_2=\cos\Omega_F$, and 
\be
 A_{B1,2} &=& -\frac{1}{2V}\int\str k_B\Lambda_{1,2}
 \Si_z(\tilde{Q}-\Si_z), \no\\
 A_{F1,2} &=& -\frac{1}{2V}\int\str k_F\Lambda_{1,2}
 \Si_z(\tilde{Q}-\Si_z), \no\\
 C_{1,2} &=& \frac{1}{2}\left[s_1A_B+s_2A_F 
 \pm(A_{B\Lambda}+A_{F\Lambda})\right], \no\\
 D_{1,2} &=& \frac{1}{2}\left[-s_1A_B+s_2A_F
 \pm (-A_{B\Lambda}+A_{F\Lambda})\right], \no\\
 A_{F,B} &=& A_{F,B 1}+A_{F,B 2}, \no\\
 A_{F,B\Lambda} &=& A_{F,B1}-A_{F,B2}.
\ee
 We neglected odd terms in $P$ as before.
 $W_1$ includes the perturbative contributions (\ref{2ptpert}) 
 and $W_2$ includes the ergodic result (\ref{2pterg}).

 In the KM domain $z_{1,2}\ll g$, 
 the expansion in $zA$ can be used.
 The integrations of the nonzero modes are calculated 
 up to second order in $1/g$.
 Introducing the renormalized mean level spacing 
 we have 
\bw
\be
 \left<\rho(\e_1)\rho(\e_2)\right> 
 &\sim& \frac{1}{\tilde{\Delta}^2}\biggl[1+\frac{a_d}{4g^2}
 +\frac{a_d}{2g^2}\left(z_1\frac{\partial}{\partial z_1}
 +z_2\frac{\partial}{\partial z_2}\right) \no\\
 & & \qquad 
 +\frac{a_d}{8g^2}\left(z_1^2\frac{\partial^2}{\partial z_1}
 +2z_1z_2\frac{\partial}{\partial z_1}\frac{\partial}{\partial z_2}
 +z_2^2\frac{\partial^2}{\partial z_2}\right) \biggr] 
 \left[\rho_1^{(0)}(z_1)\rho_1^{(0)}(z_2)-K^2(z_1,z_2)\right], 
\ee
\ew
 where $z_{1,2}=\pi\e_{1,2}/\tilde{\Delta}$, and 
 $\tilde{\Delta}$ is given by Eq.(\ref{tildeDeltaKM}).
 Subtracting the disconnected part, we derive Eq.(\ref{2ptkm}).

 The AA's domain $1\ll z_{1,2}$, $1\ll g$ is considered 
 using the asymptotic form of the Bessel function (\ref{J0asy}). 
 The details are presented in Appendix \ref{calc2}.
 From $W_1$ we obtain the first part
\be
 \left<\left<\rho(\e_1)\rho(\e_2)\right>\right>_1
 &\sim& \frac{1}{\Delta^2}\biggl[
 \frac{1}{2}\mbox{Re}\sum_{q^2\ne 0}\left(\Pi_+^2+\Pi_-^2\right) \no\\
 & & +\frac{\sin 2z_1}{2z_1}{\cal D}_1  \mbox{Im}\sum_{q^2\ne 0}
 \left(\Pi_++\Pi_-\right) \no\\
 & & +\frac{\sin 2z_2}{2z_2}{\cal D}_2  \mbox{Im}\sum_{q^2\ne 0}
 \left(\Pi_+-\Pi_{-} \right) \no\\
 & & +\frac{\cos 2(z_1+z_2)}{8z_1z_2}
 {\cal D}_1{\cal D}_2({\cal D}_+^{2}{\cal D}_-^{-2}-1) \no\\
 & &  +\frac{\cos 2(z_1-z_2)}{8z_1z_2}
 {\cal D}_1{\cal D}_2({\cal D}_-^{2}{\cal D}_+^{-2}-1)
 \biggr], \no\\ \label{rho21}
\ee
 where ${\cal D}_{1,2}={\cal D}(z_{1,2})$, 
 ${\cal D}_{\pm}={\cal D}((z_1\pm z_2)/2)$, and 
 $\Pi_\pm=\Pi(q,(\e_1\pm \e_2)/2)$.
 The first term represents the purely perturbative contribution.
 The second connected part $W_2$ is calculated 
 in the same way.
 We obtain 
\be
 \left<\left<\rho(\e_1)\rho(\e_2)\right>\right>_2
 &\sim&
 -\frac{1}{\Delta^2}\mbox{Re}\biggl\{
 \frac{1-\mbox{e}^{2i(z_1-z_2)}
 {\cal D}_1{\cal D}_2{\cal D}_-^{2}{\cal D}_+^{-2}}{2(z_1-z_2)^2} \no\\
 & & 
 +\frac{1+\mbox{e}^{2i(z_1+z_2)}
 {\cal D}_1{\cal D}_2{\cal D}_+^{2}{\cal D}_-^{-2}}{2(z_1+z_2)^2} \no\\
 & & 
 +\frac{i\left(
 \mbox{e}^{2iz_1}{\cal D}_1
 -\mbox{e}^{2iz_2}{\cal D}_2\right)}{z_1^2-z_2^2} 
 \biggr\}. \label{rho22}
\ee

 The derived expressions are written in terms of 
 the unrenormalized quantity $\Delta$ and 
 we must carry out rescaling in terms of $\tilde{\Delta}$.
 Additional contributions coming from the rescaling should cancel out  
 with terms we did not show explicitly here.
 This situation is the same as the DOS case and 
 we finally arrive at Eq.(\ref{2ptaa}). 

%%%%%%%%%%%%%%%%%%%%%%%%%%%%%%%%%%%%%%%%%%%%%%%%%%%%%%%%%%%%%%%%%%%%%%%%%%%%
%%%%%%%%%%%%%%%%%%%%%%%%%%%%%%%%%%%%%%%%%%%%%%%%%%%%%%%%%%%%%%%%%%%%%%%%%%%%
\section{The double-trace term and the DOS renormalization}
\label{double}

 In the previous section, we neglected the second term  
 in Eqs.(\ref{F1}) and (\ref{F2}).
 This double-trace term includes nonzero modes only 
 and changes the perturbative result.
 It appears only in systems with chiral symmetry 
 and we therefore concentrate on the DOS.

 At second order in $P$, we have instead of Eq.(\ref{F10})
\be
 F_1^{(0)} &=& \frac{\pi D}{\Delta V}\int\str (\nabla P)^2
 +\frac{\pi D_1}{4\Delta V}\int\left(\str \nabla P\Si_x\right)^2 \no\\
 & & -\frac{i\pi\e}{\Delta V}\int\str P^2.
 \label{F10d1}
\ee
 The presence of the second term 
 modifies the contraction rule as Eq.(\ref{cont1d1}). 
 In this case perturbation theory is formulated by 
 expansions in $\Pi$, Eq.(\ref{pi}), and  
\be
 \Pi_2(q,\e) = \frac{\pi D_1q^2}{\Delta}\Pi^2(q,\e). \label{pi2}
\ee
 The corresponding expansion parameters are 
 $1/g\propto 1/D$ and $g_1/g^2\propto D_1/D^2$.

 The perturbative expansion gives the DOS 
\be
 \left<\rho(\e)\right> &=& \frac{1}{\Delta}\mbox{Re}
 \biggl[1+\sum_q\Pi_2(q,\e)
 +\frac{1}{2}\Bigl(\sum_q\Pi(q,\e)\Bigr)^2 \no\\
 & & +\frac{1}{2}\Bigl(\sum_q\Pi_2(q,\e)\Bigr)^2 +\cdots
 \biggr]. \label{dospert2}
\ee
 The new propagator $\Pi_2$ contributes to the DOS at one-loop order.
 The renormalized mean level spacing is defined as 
 the inverse of Eq.(\ref{dospert2}) excluding zero-mode contributions.
 Actually this result was derived by Gade using 
 the renormalization group method \cite{Gade}.
 In our model (\ref{F1}), following the calculation in Ref.\cite{Efetov}, 
 we can obtain the same renormalization group equations 
 at one-loop order as 
\be
 \beta_b = -\frac{db}{d\mu} = \e b, \quad
 \beta_c = -\frac{dc}{d\mu} = \e c+c^2, \quad
 \zeta = \frac{b^2}{c},
\ee
 where $b\sim 1/g$ and $c\sim 1/g_1$. 
 We used $\e$ expansion, $\e=d-2$, 
 and $\mu$ is the renormalization scale.
 $\beta_{b,c}$ are the beta functions for $b$ and $c$. 
 $\zeta$ is the zeta function for 
 the wavefunction renormalization and 
 corresponds to the result in Eq.(\ref{dospert2}).
 These equations imply a divergence of the DOS 
 and delocalization of eigenstates in two dimension. 
 Thus the presence of the double-trace term changes 
 the behavior of the DOS significantly.
 We note that the renormalization procedure produces the double-trace 
 term even if we start the analysis from a model without that term.
 The quantum effect in two dimension increases 
 the coupling constant $c$.

 We consider the scaled DOS $\rho_1(z)$
 to examine how the double-trace term contributes to the result.
 The perturbative result (\ref{dospert2}) is renormalized to 
 the mean level spacing.
 In a similar way as the calculation in the previous section
 we find in the KM domain $z\ll g$ 
\be
 \rho_1(z) &\sim&
 \biggl\{1+\biggl[
 \frac{a_d}{8g^2}
 +\frac{a_d}{16}\left(\frac{g_1}{g^2}\right)^2
 \biggr] \no\\
 & & \times
 \left(2z\frac{d}{dz}
 +z^2\frac{d^2}{dz^2}\right)\biggr\} \rho_1^{(0)}(z). \label{doskm2}
\ee
 In the AA domain $1\ll z$, 
 the spectral determinant is modified as Eq.(\ref{sdetd1}).
 Subtracting the renormalization effect, we obtain 
\be
 {\cal D}(z) &\sim& 
 \prod_{q\ge 0,q^2\ne 0}
 \frac{(Dq^2)^2}{(Dq^2)^2+\e^2} \no\\
 & & \times
 \biggl[1-8z^2\sum_{q\ge 0,q^2\ne 0}\left|\Pi_2(q,\e)\right|^2\biggr], 
\ee
 which is consistent with Eq.(\ref{doskm2}).

 Finally we mention the Jacobian contribution in Eq.(\ref{jacobian}).
 It includes a term second order in $P$ and 
 changes the contraction rules.
 Since this term is similar to the last term in Eq.(\ref{F10d1})
 it can be easily incorporated
 into the contraction rules by the replacement 
\be
 \Pi_2(q,\e) \to \Pi_2(q,\e)-\Pi^{2}(q,\e).
\ee
 Thus this Jacobian contribution is always subleading 
 compared to the propagator $\Pi_2$.
 We also note that this contributes only to $\tilde{\Delta}$ and not to 
 the scaled DOS $\rho_1(z)$ in our approximation.

%%%%%%%%%%%%%%%%%%%%%%%%%%%%%%%%%%%%%%%%%%%%%%%%%%%%%%%%%%%%%%%%%%%%%%%%%%%%
%%%%%%%%%%%%%%%%%%%%%%%%%%%%%%%%%%%%%%%%%%%%%%%%%%%%%%%%%%%%%%%%%%%%%%%%%%%%
\section{Discussions and Conclusions}
\label{conc}

% conclusions

 We have studied disordered systems with chiral unitary symmetry.
 Using a chiral symmetric random matrix model 
 we derived the nonlinear $\sigma$ models (\ref{F1}) and (\ref{F2}).
 We demonstrated that they are equivalent to 
 related chiral symmetric models. 
 Using the $\sigma$ models, 
 we calculated the level correlation functions.
 We exploited the nonperturbative methods 
 developed by Kravtsov and Mirlin, and Andreev and Altshuler
 for the traditional classes.

% comparison of models

 The equivalence of models shows the universality of disordered systems.
 Our derived $\sigma$ models are applicable to models 
 treated in Refs.\cite{Gade,rfm,th}. 
 The double-trace term was not derived in Ref.\cite{th}.
 This is because the massive mode integration 
 was not considered carefully.

% level correlation functions

 For the calculation of the DOS and TLCF, 
 we stressed the need for the renormalization of the mean level spacing. 
 This renormalization is absent in traditional nonchiral systems.
 After separating the renormalization effect, we found the results 
 (\ref{doskm}) and (\ref{2ptkm}) in the KM domain, and 
 (\ref{dosaa}) and (\ref{2ptaa}) in the AA domain.
 It is interesting to note that 
 the results in the AA domain are expressed 
 using the spectral determinant.
 It contributes to oscillating terms only, 
 in a similar way as for the traditional classes.
 Thus we conclude that 
 the singularity of the form factor at the Heisenberg time
 is washed out due to finite-$g$ effects \cite{AA}.

% comparison of the result for the level correlation functions

 Our formulation of the perturbative and nonperturbative
 calculations can be useful 
 not only for the level correlation functions
 but also for the conductance and other quantities.
 In the present work 
 we concentrated on the level correlation functions. 
 In Ref.\cite{th}, the same quantities were calculated perturbatively.
 The different result obtained there is 
 due to another parametrization of the $Q$ matrix. 
 Additional contributions 
 coming from the integration measure would give the correct result.
 In Ref.\cite{TI}, 
 the DOS in the KM domain was calculated 
 from the model derived in Ref.\cite{th}.
 The result was scaled in terms of the bare mean level spacing $\Delta$, 
 and the renormalized mean level spacing $\tilde{\Delta}$ 
 was not introduced. 
 This leads us to a different conclusion on level statistics 
 as we mention below.
 
% dos renormalization

 Let us discuss the importance of introducing  
 the renormalized mean level spacing.
 There are numerous works on the behavior of the DOS 
 at the origin $\e=0$.
 The main question is whether the DOS diverges or not, 
 and analytically it has been considered using 
 perturbation theory at weak disorder $g\gg 1$.
 On the other hand, chiral RMT, 
 which corresponds to the model at $g=\infty$, 
 predicts the vanishing of the DOS at the origin $z=0$.
 This is not a contradiction and 
 indicates the importance of scaling.
 The macroscopic behavior is determined by the nonzero modes
 and a divergence of $1/\tilde{\Delta}$ was reported in Ref.\cite{Gade}. 
 The zero mode has nothing to do with this behavior 
 and determines the universal behavior at the microscopic scale.
 It can be seen by scaling the energy variable $\e$ 
 in terms of the mean level spacing.

 Generally speaking, 
 the behavior at the macroscopic scale depends on the model.
 From a field theoretical point of view, 
 the divergence can be renormalized to the mean level spacing and 
 a definite conclusion as to whether it is a real divergence or not 
 can be obtained by referring to 
 other approaches such as numerical simulations.
 Our result relies on perturbation theory 
 and the divergence may be cut off somewhere before the origin.
 This crossover to the universal microscopic domain 
 is highly nonperturbative.
 Since a high resolution is required, 
 it may be hard to see such a crossover numerically. 

 From the viewpoint of level statistics, 
 the DOS must be scaled (renormalized) to unity at all energies
 to find the universality.
 This unfolding procedure cannot be applied 
 to the present chiral case 
 because the DOS itself has the universal fine structure 
 (oscillations due to level repulsion) at the origin.
 For this reason, we use $\Delta$, 
 the (inverse) DOS at $z=\infty$ ($\e=0$), 
 for scaling in the ergodic regime.
 It is modified by finite-$g$ effects  
 and we use $\tilde{\Delta}$ to see the microscopic domain closely.

 Thus using the renormalized mean level spacing, 
 we can separate problems at both scales. 
 The effects of nonzero modes (finite-$g$ effect) 
 cannot be scaled out completely in the microscopic domain and 
 deviations from the universal behavior are obtained 
 as we have shown in the present work.
 Such an example can be found in Ref.\cite{GT}.
 The generalized random matrix model was used there and 
 it was found that the quantity $\tilde{\Delta}$ is different from 
 our result.
 However, after scaling in terms of 
 the nonuniversal quantity $\tilde{\Delta}$, 
 we can find the complete agreement up to finite-$g$ corrections.
 This demonstration of ``universal deviation'' 
 justifies the introduction of $\tilde{\Delta}$.

% double-trace term

 The double trace term contribution is small at the classical level 
 because the coupling constant is small compared with 
 that in the diffusion propagator.
 However, quantum effects affect this coupling and 
 the contribution becomes important in some cases.
 It significantly affects the DOS renormalization and 
 a diverging DOS was found in Ref.\cite{Gade}.
 Concerning level statistics, 
 this term modifies the spectral determinant as Eq.(\ref{sdetd1}).

% orthogonal and symplectic

 Our calculation is only for the chiral unitary class. 
 The other chiral classes, chiral orthogonal and symplectic, 
 can be calculated in the same way.
 The problem is that the proper parametrization of the zero mode 
 has not been found.
 However the KM's domain can be considered without knowing 
 the zero mode parametrization as was done in Ref.\cite{TI}.
 The obtained result is valid only at first order in $1/g$.
 Repeating the same calculation up to the next order 
 and introducing the renormalized mean level spacing, 
 we found the same form as Eq.(\ref{doskm}).
 The coefficient of the second term in Eq.(\ref{doskm}) 
 is changed but with the same sign for all the classes. 
 This result also holds for the TLCF (\ref{2ptkm}). 
 We thus obtain the same conclusion as KM, namely, 
 the weakening of level repulsion 
 [This can be seen, e.g., in Eq.(\ref{lr})].
 The authors in Ref.\cite{TI} drew a different conclusion 
 by looking at the first order correction to the mean level spacing.
 It is renormalized to the mean level spacing and 
 should be applied to the DOS behavior and not to level repulsion.

% critical statistics

 As an interesting application, 
 we mention a related work in Ref.\cite{GV}.
 For traditional nonchiral systems, 
 the authors in Ref.\cite{KT} pointed out that 
 the AA's result is related to the Calogero-Sutherland model 
 at finite temperature.
 It is shown in Ref.\cite{GV2} that this model is equal to 
 the generalized random matrix model proposed in Ref.\cite{MNS}.
 In this problem the nonlinear $\sigma$ model is modified 
 due to power-law correlations of random matrices \cite{PRBM} 
 and the diffusion propagator and spectral determinant are modified.
 As a result agreement with the result in Ref.\cite{MNS} was found 
 and a conjecture to more general cases was made.
 We expect this holds also for chiral systems and 
 the result is presented in Ref.\cite{GT}.

% wavefunction statistics

 Another future problem is the wavefunction statistics.
 For traditional classes, 
 it was considered in Ref.\cite{FM} using the KM method.
 It will be interesting to see how this result is modified 
 in the chiral symmetric case.

%%%%%%%%%%%%%%%%%%%%%%%%%%%
\section*{Acknowledgments}

 The author is grateful to 
 A.M. Garc\'{\i}a-Garc\'{\i}a, S. Iida, and G. Schwiete 
 for useful discussions and reading the manuscript.
 The financial support by the SFB/Transregio 12 is acknowledged.

%%%%%%%%%%%%%%%%%%%%%%%%%%%%%%%%%%%%%%%%%%%%%%%%%%%%%%%%%%%%%%%%%%%%%%%%%%%%
%%%%%%%%%%%%%%%%%%%%%%%%%%%%%%%%%%%%%%%%%%%%%%%%%%%%%%%%%%%%%%%%%%%%%%%%%%%%
\appendix
%%%%%%%%%%%%%%%%%%%%%%%%%%%%%%%%%%%%%%%%%%%%%%%%%%%%%%%%%%%%%%%%%%%%%%%%%%%%
\section{Contraction rules}
\label{cont}
%%%%%%%%%%%%%%%%%%%%%%%%%%%%%%%%%%%%%%%%%%%%%%%%%%%%%%%%%%%%%%%%%%%%%%%%%%%%
\subsection{Calculation for $F_1$}

 Consider the functional 
\be
 F_1^{(2)} &=& \frac{\pi D}{\Delta V}\int\str (\nabla P)^2 
 -\frac{iz}{V}\lambda_F\int\str k_FP^2 \no\\
 & & -\frac{iz}{V}\lambda_B\int\str k_BP^2, \label{F10k}
\ee
 where $k_{F,B}=(1\pm k)/2$ and $z=\pi \e/\Delta$.
 The $P$ matrix is a 4$\times$4 supermatrix 
 including nonzero modes and is given by Eq.(\ref{P}).
 Since this functional breaks supersymmetry
 for $\lambda_F\ne\lambda_B$, the function
 ${\cal D}_1(z,\lambda_B,\lambda_F)=\int{\cal D}\tilde{Q}\exp(-F_1^{(2)})$ 
 is not normalized to unity.
 We calculate this function and derive the contraction rules.

 Using the explicit parametrization (\ref{P}), 
 we write $F_1^{(2)}$ as 
\be
 F_1^{(2)} &=& \sum_{q\ne 0} 
 \left(\ba{cccc} -\rho & \sigma & a & b \ea \right)(-q) 
 G^{-1}
 \left(\ba{c} \sigma \\ \rho \\ a \\ b \ea \right)(q) \no\\
 &\equiv& \sum_{q\ne 0} 
 \bar{\psi}(q)G^{-1}\psi(q), \no\\
 G &=& \mbox{diag}(
 \Pi(q,\e\lambda_+), \Pi(q,\e\lambda_+),
 \Pi(q,\e\lambda_F), \Pi(q,\e\lambda_B)), \no\\ \label{G}
\ee
 where $\lambda_+=(\lambda_B+\lambda_F)/2$ and 
 the diffusion propagator is given by Eq.(\ref{pi}).
 Then the functional integral is given by
\be 
 & & {\cal D}_1(z,\lambda_B,\lambda_F) =
 \int {\cal D}(\bar{\psi},\psi)\mbox{e}^{-F_1^{(2)}} \no\\
 &=&  \prod_{q^2\ne 0} 
 \left[\frac{\Pi(q,\e\lambda_B)\Pi(q,\e\lambda_F)}
 {\Pi^2(q,\e\lambda_+)}\right]^{1/2} \no\\
 &=& \prod_{q\ge 0,q^2\ne 0} 
 \frac{(Dq^2-i\e\lambda_+)^2}{(Dq^2-i\e\lambda_B)(Dq^2-i\e\lambda_F)},
\ee
\be
 \int {\cal D}(\bar{\psi},\psi)\psi\bar{\psi}\mbox{e}^{-F_1^{(2)}} 
 = \frac{1}{2}{\cal D}_1(z,\lambda_B,\lambda_F)G.
\ee
 Since $\psi(q)$ and $\psi(-q)$ are not independent of each other, 
 the square root appears in ${\cal D}$.
 We used the periodic boundary condition and $q_i=2\pi n_i/L$, 
 $n_i$ is integer.

 Using the result we obtain the contraction rules for the matrix $P$ as
\bw
\be
 \left<\str AP(x)BP(y)\right> &=& 
 \frac{1}{4}
 \sum_{i,j=F,B}\Pi\left(x-y,\e\frac{\lambda_i+\lambda_j}{2}\right)
 ( \str k_i A \str k_j B-\str k_i A\Si_z \str k_j B\Si_z \no\\
 & & 
 +\str k_i A\Si_x \str k_j B\Si_x
 -\str k_i A\Si_y \str k_j B\Si_y ), \no\\
 \left<\str AP(x)\,\str BP(y)\right> &=& 
 \frac{1}{4}
 \sum_{i,j=F,B} \Pi\left(x-y,\e\frac{\lambda_i+\lambda_j}{2}\right) \no\\ 
 & & \times
 \str \left( k_i A k_j B-k_i A \Si_z k_j B \Si_z 
 +k_i A\Si_x k_j B\Si_x-k_i A\Si_y k_j B\Si_y \right),
\ee
\ew
 where $A$ and $B$ are arbitrary supermatrices and 
 $\left<\cdots\right> = {\cal D}_1^{-1}(z,\lambda_B,\lambda_F)
 \int{\cal D}\tilde{Q}(\cdots)\exp(-F_1^{(2)})$. 

 We are mainly interested in the case 
 $(\lambda_B, \lambda_F)=(1, 1)$, and $(1, -1)$. 
 The first case $(1,1)$ corresponds to standard perturbation theory.
 The free energy does not break supersymmetry 
 and we find ${\cal D}(z,1,1)=1$ and 
\be
 \left<\str AP(x)BP(y)\right> &=& \frac{1}{4}\Pi(x-y,\e) \no\\
 & & \times(\str A \str B-\str A\Si_z\str B\Si_z \no\\
 & & +\str A\Si_x \str B\Si_x \no\\
 & & -\str A\Si_y \str B\Si_y), \no\\
 \left<\str AP(x)\str BP(y)\right> &=& \frac{1}{4}\Pi(x-y,\e) \no\\ 
 & & \times\str ( AB-A\Si_zB\Si_z \no\\
 & & +A\Si_xB\Si_x-A\Si_yB\Si_y). \label{cont1}
\ee
 In the case $(1, -1)$, supersymmetry is broken 
 and this is used for calculations in the AA's domain $1\ll z$.
 The function ${\cal D}$ is given by 
\be
 {\cal D}_1(z,1,-1) &=& \prod_{q\ge 0,q^2\ne 0}
 \frac{(Dq^2)^2}{(Dq^2)^2+\e^2} \no\\
 &=& \prod_{n\ge 0,n^2\ne 0}
 \frac{g^2(4\pi^2 n^2)^2}{g^2(4\pi^2 n^2)^2+z^2}. \label{sdet0}
\ee

%%%%%%%%%%%%%%%%%%%%%%%%%%%%%%%%%%%%%%%%%%%%%%%%%%%%%%%%%%%%%%%%%%%%%%%%%%%%
\subsection{Effect of the double-trace term}

 We consider the effect of the double-trace term.
 The second term of Eq.(\ref{F10d1}) is included in Eq.(\ref{F10k}).
 In this case, the matrix $G$ in Eq.(\ref{G}) is replaced by
\be
 G(q) &=& \mbox{diag}
 \left(\Pi_+, \Pi_+, C\Pi_F, C\Pi_B \right)  \no\\
 & & -\frac{\pi D_1q^2}{\Delta}\Pi_F\Pi_B
 C\left(\ba{cccc}
 0 & 0 & 0 & 0 \\ 0 & 0 & 0 & 0 \\
 0 & 0 & 1 & -i \\ 0 & 0 & -i & -1
 \ea\right), \no\\
 \Pi_{F,B,+} &=& \frac{\Delta}{2\pi}\frac{1}{Dq^2-i\e\lambda_{F,B,+}}, \no\\
 C &=& \frac{1}{1+\frac{\pi D_1q^2}{\Delta}(\Pi_F-\Pi_B)}.
\ee
 As a result, ${\cal D}$ and contraction rules
 are modified in the following way.
 The contraction for $(\lambda_B,\lambda_F)=(1,1)$ is expressed as
\bw
\be
 \left<\str AP(x)BP(y)\right> &=& 
 \frac{1}{4}\Pi(x-y,\e)
 \left(\str A \str B-\str A\Si_z\str B\Si_z 
 +\str A\Si_x \str B\Si_x-\str A\Si_y \str B\Si_y \right)  \no\\
 & & -\frac{1}{2}\Pi_2(x-y,\e)\str A\Si_x B\Si_x, 
 \no\\
 \left<\str AP(x)\str BP(y)\right> &=& \frac{1}{4}\Pi(x-y,\e)
 \str \left(AB-A\Si_zB\Si_z+A\Si_xB\Si_x-A\Si_yB\Si_y\right) \no\\
 & & -\frac{1}{2}\Pi_2(x-y,\e)
 \str A\Si_x\str B\Si_x, \label{cont1d1}
\ee
\ew
 where the propagator $\Pi_2$ in momentum space is given by (\ref{pi2}).
 For $(\lambda_B,\lambda_F)=(1,-1)$, (\ref{sdet0}) is replaced by
\be
 {\cal D}_1(z,1,-1) = \prod_{q\ge 0,q^2\ne 0}
 \frac{(Dq^2)^2}{(Dq^2)^2-i\e D_1q^2+\e^2}. \label{sdetd1}
\ee

%%%%%%%%%%%%%%%%%%%%%%%%%%%%%%%%%%%%%%%%%%%%%%%%%%%%%%%%%%%%%%%%%%%%%%%%%%%%
\subsection{Calculation for $F_2$}

 Consider 
\be
 & & F_2^{(2)}(z_1,z_2,z_3,z_4) =
 \frac{\pi D}{\Delta V}\int\str (\nabla P)^2 \no\\
 & & -\frac{iz_1}{V}\int\str k_F\Lambda_1 P^2 
 -\frac{iz_2}{V}\int\str k_B\Lambda_1 P^2 \no\\
 & & -\frac{iz_3}{V}\int\str k_F\Lambda_2 P^2 
 -\frac{iz_4}{V}\int\str k_B\Lambda_2 P^2, \label{F20k}
\ee
 where $\Lambda_{1,2}=(1\pm\Lambda)/2$ and 
 $z_{1,2,3,4}=\pi\e_{1,2,3,4}/\Delta$.
 The $P$ matrix is an 8$\times$8 supermatrix 
 and is parametrized as Eq.(\ref{P2}).
 This case is considered in the same way as the case of $F_1^{(2)}$.
 We neglect the double-trace term contribution for simplicity.
 The result is expressed for the functional integral as 
\bw
\be
 {\cal D}_2(z_1,z_2,z_3,z_4) &=& \int {\cal D}\tilde{Q}
 \mbox{e}^{-F_2^{(2)}(z_1,z_2,z_3,z_4)}
 = {\cal D}_{\rm ch}(z_1,z_2)
 {\cal D}_{\rm ch}(z_3,z_4)
 {\cal D}_{\rm u}(z_1,z_2,z_3,z_4), 
 \no\\
 {\cal D}_{\rm ch}(z_1,z_2) &=& \prod_{q\ge 0,q^2\ne 0}
 \frac{\left[Dq^2-\frac{i}{2}(\e_1+\e_2)\right]^2}{(Dq^2-i\e_1)(Dq^2-i\e_2)},
 \no\\
 {\cal D}_{\rm u}(z_1,z_2,z_3,z_4) &=& \prod_{q\ge 0,q^2\ne 0}
 \frac{Dq^2-\frac{i}{2}(\e_1+\e_4)}
 {Dq^2-\frac{i}{2}(\e_1+\e_3)}
 \frac{Dq^2-\frac{i}{2}(\e_2+\e_3)}
 {Dq^2-\frac{i}{2}(\e_2+\e_4)}. \label{calD2}
\ee
 For example, 
 ${\cal D}_2(-z_1,z_1,z_2,z_2)={\cal D}_1(z_1)$,
 ${\cal D}_2(z_1,z_1,-z_2,z_2)={\cal D}_1(z_2)$, and  
 ${\cal D}_2(-z_1,z_1,-z_2,z_2)={\cal D}_1(z_1){\cal D}_1(z_2)
 {\cal D}_1^{2}[(z_1+z_2)/2]{\cal D}_1^{-2}[(z_1-z_2)/2]$.
 The contraction rule is given by
\be
 & & \left<\str AP(x)BP(y)\right> =
 \frac{1}{4}\sum_{\alpha,\beta=F,B} \sum_{i,j=1,2}
 \Pi_{i\alpha j\beta}\left(x-y\right) \no\\
 & & \times
 \left(\str k_\alpha\Lambda_i A \str k_\beta\Lambda_j B
 -\str k_\alpha\Lambda_i\Si_z A\str k_\beta\Lambda_j\Si_z B
 +\str k_\alpha\Lambda_i\Si_x A \str k_\beta\Lambda_j\Si_x B
 -\str k_\alpha\Lambda_i\Si_y A \str k_\beta\Lambda_j\Si_y B \right), \no\\
 & & \left<\str AP(x)\,\str BP(y)\right> = 
 \frac{1}{4}\sum_{\alpha,\beta=F,B} \sum_{i,j=1,2}
 \Pi_{i\alpha j\beta}(x-y) \no\\
 & & \times
 \str \left( k_\alpha\Lambda_i A k_\beta\Lambda_j B
 -k_\alpha\Lambda_i\Si_z A k_\beta\Lambda_j\Si_z B
 +k_\alpha\Lambda_i\Si_x A k_\beta\Lambda_j\Si_x B
 -k_\alpha\Lambda_i\Si_y A k_\beta\Lambda_j\Si_y B \right), 
\ee
 where $\Pi_{i\alpha j\beta}(x)=
 \Pi(x,(\lambda_{i\alpha}+\lambda_{j\beta})/2)$, 
 and $\lambda_{\rm 1F}=\e_1$, $\lambda_{\rm 1B}=\e_2$,  
 $\lambda_{\rm 2F}=\e_3$, $\lambda_{\rm 2B}=\e_4$.
 The case $(z_1,z_2,z_3,z_4)\to(z_1,z_1,z_2,z_2)$
 corresponds to standard perturbation and we find
\be
 \left<\str AP(x)BP(y)\right> &=& 
 \frac{1}{4}\sum_{i,j=1,2}\Pi\left(x-y,\frac{\e_i+\e_j}{2}\right)
 (\str A\Lambda_i\str B\Lambda_j
 -\str A\Lambda_i\Si_z\str B\Lambda_j\Si_z \no\\
 & & 
 +\str A\Lambda_i\Si_x\str B\Lambda_j\Si_x
 -\str A\Lambda_i\Si_y\str B\Lambda_j\Si_y), 
 \no\\
 \left<\str AP(x)\str BP(y)\right> &=& 
 \frac{1}{4}\sum_{i,j=1,2}\Pi\left(x-y,\frac{\e_i+\e_j}{2}\right)
 \str (A\Lambda_iB\Lambda_j-A\Lambda_i\Si_z B\Lambda_j\Si_z \no\\
 & & 
 +A\Lambda_i\Si_x B\Lambda_j\Si_x
 -A\Lambda_i\Si_yB\Lambda_j\Si_y). \label{cont2}
\ee
\ew

%%%%%%%%%%%%%%%%%%%%%%%%%%%%%%%%%%%%%%%%%%%%%%%%%%%%%%%%%%%%%%%%%%%%%%%%%%%%
%%%%%%%%%%%%%%%%%%%%%%%%%%%%%%%%%%%%%%%%%%%%%%%%%%%%%%%%%%%%%%%%%%%%%%%%%%%%
\section{Calculation of the two-level correlation function}
\label{calc2}

%%%%%%%%%%%%%%%%%%%%%%%%%%%%%%%%%%%%%%%%%%%%%%%%%%%%%%%%%%%%%%%%%%%%%%%%%%%%
\subsection{Zero mode integration}

 In this section we derive Eq.(\ref{W12}) 
 by integrating the zero-mode variables 
 of the nonperturbative parametrization (\ref{Qnonp}).
 As before, the parametrization is slightly modified as 
\be
 Q(x) &=& U_{\rm u}T_{\rm ch}T_{\rm u}^{(0)}\tilde{Q}(x)
 \bar{T}_{\rm u}^{(0)}\bar{T}_{\rm ch}\bar{U}_{\rm u},
\ee
 to eliminate the grassmann variables of unitary part in $F_2$.
 For the pre-exponential term,
 dependence of the grassmann variables on $U_{\rm u}$ is 
 explicitly written as 
\be
 & & \str k\Lambda_1\Sigma_z Q(x) \str k\Lambda_2\Sigma_z Q(x) \no\\
 &\to& 
 \str k\Lambda_1\Sigma_z T_{\rm ch}\tilde{Q}(x)\bar{T}_{\rm ch}
 \str k\Lambda_2\Sigma_z T_{\rm ch}\tilde{Q}(x)\bar{T}_{\rm ch} \no\\
 & & 
 -4\xi\xi^*\eta\eta^*
 \str\Lambda_1\Sigma_z 
 T_{\rm ch}T_{\rm u}^{(0)}\tilde{Q}(x)
 \bar{T}_{\rm u}^{(0)}\bar{T}_{\rm ch} \no\\
 & & \times\str\Lambda_2\Sigma_z 
 T_{\rm ch}T_{\rm u}^{(0)}\tilde{Q}(x)\bar{T}_{\rm u}^{(0)}\bar{T}_{\rm ch}.
\ee
 The neglected terms do not contribute to integration 
 of the Grassmann variables.
 The first term does not include the grassmann variables $\xi$ and $\eta$.
 We can set $T_{\rm u}^{(0)}=1$ and have
\bw
\be
 W_1(\e_1,\e_2)
 &=& \frac{1}{16\Delta^2V^2}\int {\cal D}Q
 \left[\int_x \str k\Lambda_1\Sigma_z Q(x)\right] 
 \left[\int_y \str k\Lambda_2\Sigma_z Q(y)\right] 
 \mbox{e}^{-F_2[Q]}.
\ee
 where $Q=T_{\rm ch}\tilde{Q}\bar{T}_{\rm ch}$.
 It still includes the zero-mode variables 
 of the chiral part $T_{\rm ch}$.
 Since the chiral part parametrization is the same as that 
 of the DOS, the calculation can be done as in Sec.\ref{dos}.
 As a result $W_1(\e_1,\e_2)$ in Eq.(\ref{W12}) is obtained.
 It includes a perturbative part and 
 connected and disconnected parts.

 Next we consider the second contribution which 
 includes only the connected part. 
 It is obtained by integrations of $\xi$ and $\eta$ as 
\be
 W_2(\e_1,\e_2)
 &=& \frac{1}{\Delta^2}\int 
 ds_1ds_2\frac{d\varphi_B}{2\pi}\frac{d\varphi_F}{2\pi} 
 \frac{4s_1s_2}{(s_1^2-s_2^2)^2}
 I(z_{1,2},s_{1,2},\varphi_{B,F}), \no\\
 I(z_{1,2},s_{1,2},\varphi_{B,F}) 
 &=& -\frac{\partial}{\partial z_1}\frac{\partial}{\partial z_2}
 \int{\cal D}\tilde{Q}{\cal D}Q_{\rm ch} \ \mbox{e}^{-F[Q]},
\ee
 where $Q(x)=T_{\rm ch}T_{\rm u}^{(0)}
 \tilde{Q}(x)\bar{T}_{\rm u}^{(0)}\bar{T}_{\rm ch}$, 
 $Q_{\rm ch}=T_{\rm ch}\Sigma_z\bar{T}_{\rm ch}$,  
 and $z_{1,2}=\pi\e_{1,2}/\Delta$.
 We examine $\str \hat{\e} \Sigma_z Q(x)$
 to integrate out variables in $Q_{\rm ch}$.
 The expression is simplified if we apply 
 the saddle point approximation we use in the following.
 At the saddle-point we have 
 $\sin\hat{\theta}=0$ and $\sin\hat{\Omega}=0$.
 This approximation leads to the reduction
\be
 \str \hat{\e} \Sigma_z Q(x) 
 &\to& \e_1\str\frac{\cos\hat{\Omega}+\Lambda}{2}
 \left[\cos\hat{\theta}_1+(\cosh\theta_{1B}-\cos\theta_{1F})
 \rho_1\sigma_1\right]
 \Sigma_z\tilde{Q}(x) \no\\
 & & +\e_2\str\frac{\cos\hat{\Omega}-\Lambda}{2}
 \left[\cos\hat{\theta}_2+(\cosh\theta_{2B}-\cos\theta_{2F})
 \rho_2\sigma_2\right]
 \Sigma_z\tilde{Q}(x).
\ee
 Again we stress that this approximation is justified 
 at second order in $1/g$.
 Substituting this expression, we have 
\be
 I(z,s,\varphi) &=& -\int{\cal D}\tilde{Q}\mbox{e}^{-F_{\rm kin}}
 I_1(z_1,s)I_2(z_2,s), \\
 I_{i=1,2}(z,s) 
 &=& \frac{\partial}{\partial z}\int{\cal D}Q_{\rm ch}
 \biggl[1-\frac{iz}{2V}\int_x\str\frac{\cos\hat{\Omega}\pm\Lambda}{2}
 (\cosh\theta_{iB}-\cos\theta_{iF})
 \rho_i\sigma_i\Sigma_z\tilde{Q}(x) \biggr] \no\\
 & & \qquad\qquad\times
 \exp\biggl[-\frac{iz}{2V}\int_x\str\frac{\cos\hat{\Omega}\pm\Lambda}{2}
 \cos\hat{\theta}_i\Sigma_z\tilde{Q}(x) \biggr],
\ee
 where $F_{\rm kin}$ is the kinetic part in $F_2$.
 The variable $\varphi$ is not included in 
 the integrand in our approximation.
 Integrations of the remaining zero-mode variables
 are carried out and we find 
\be
 I_i(z,s) &=& i(s_1-s_2+C_i)\Biggl\{\mbox{e}^{iz(s_1-s_2)+izC_i}
 +\frac{1}{4\pi}\int d\theta_{iB}d\theta_{iF}
 \frac{\lambda_{iB}\lambda_{iF}-1}{\lambda_{iB}-\lambda_{iF}} \no\\
 & & \qquad\times
 \biggl[1+iz\left(s_1\lambda_{iB}-s_2\lambda_{iF}
 +\frac{1}{2}(C_i-D_i)\lambda_{iB}
 +\frac{1}{2}(C_i+D_i)\lambda_{iF}\right)\biggl] \no\\
 & & \qquad\times
 \exp\biggl[
 iz\left(s_1\lambda_{iB}-s_2\lambda_{iF}+\frac{1}{2}(C_i-D_i)\lambda_{iB}
 +\frac{1}{2}(C_i+D_i)\lambda_{iF}\right)\biggr]
 \Biggr\} \no\\
 &=& \frac{i\pi z}{2}(s_1-s_2+C_i)(s_1+s_2-D_i)
 J_0\left(zs_1+z\frac{C_i-D_i}{2}\right)
 J_0\left(zs_2-z\frac{C_i+D_i}{2}\right).
\ee
 where $\lambda_{iF}=\cos\theta_{iF}$ and $\lambda_{iB}=\cosh\theta_{iB}$.
 This result yields $W_2$ in Eq.(\ref{W12}).

%%%%%%%%%%%%%%%%%%%%%%%%%%%%%%%%%%%%%%%%%%%%%%%%%%%%%%%%%%%%%%%%%%%%%%%%%%%%
\subsection{AA's domain}

 We consider Eq.(\ref{W12}) in the AA domain $1\ll z_{1,2}$ 
 using the asymptotic form of the Bessel function (\ref{J0asy}). 
 For $W_1$, 
\be
 W_1(\e_1,\e_2)
 &\sim& \frac{1}{\Delta^2}\biggl<
 \left[f_1(z_1)\mbox{e}^{iz_1(A_{B1}+A_{F1})} 
 +g_1(z_1)\mbox{e}^{2iz_1+iz_1(A_{B1}-A_{F1})}\right] \no\\
 & & \times
 \left[f_2(z_2)\mbox{e}^{iz_2(A_{B2}+A_{F2})} 
 +g_2(z_2)\mbox{e}^{2iz_2+iz_2(A_{B2}-A_{F2})}\right]
 \biggr>_{\rm kin}, \no \\
 &=& \frac{1}{\Delta^2}\int{\cal D}\tilde{Q}
 \Bigl[\mbox{e}^{-F(z_1,z_1,z_2,z_2)}f_1(z_1)f_2(z_2)
 +\mbox{e}^{2iz_1-F(-z_1,z_1,z_2,z_2)}g_1(z_1)g_2(z_2) \no\\
 & & +\mbox{e}^{2iz_2-F(z_1,z_1,-z_2,z_2)}f_1(z_1)g_2(z_2)
 +\mbox{e}^{2i(z_1+z_2)-F(-z_1,z_1,-z_2,z_2)}g_1(z_1)g_2(z_2)
 \Bigr], \label{W1AA}
\ee
\ew
 where 
\be
 f_i(z) &=& 1+\frac{1}{8z^2}+\frac{1}{2}(A_{Bi}-A_{Fi})+\cdots, \no\\
 g_i(z) &=& -\frac{1}{2z}+\frac{i}{8z^2}+\cdots, 
\ee
 and $F$ is the supersymmetry breaking functional 
\be
 & & F(z_1,z_2,z_3,z_4) =  
 \frac{\pi D}{4\Delta V}\int\str (\nabla \tilde{Q})^2 \no\\
 & & \quad
 +\frac{i}{2V}\int\str (z_1k_F+z_2k_B)
 \Lambda_1\Si_z(\tilde{Q}-\Si_z) \no\\
 & & \quad
 +\frac{i}{2V}\int\str (z_3k_F+z_4k_B)
 \Lambda_2\Si_z(\tilde{Q}-\Si_z).
\ee
 The condition $z_1=z_2$ and $z_3=z_4$ recovers supersymmetry.
 As before we expand the nonzero modes $\tilde{Q}$
 in terms of the $P$ matrix and 
 use the contraction rules derived in Appendix \ref{cont}.
 The first term in Eq.(\ref{W1AA}) does not break supersymmetry 
 [$F(z_1,z_1,z_2,z_2)=F_2$] 
 and is nothing but the purely perturbative contribution.
 Its connected part gives the first term in Eq.(\ref{rho21}).
 The second (third) term in Eq.(\ref{W1AA}) gives 
 the second (third) term in Eq.(\ref{rho21}).
 For the leading order contribution, we use 
\be
 \int{\cal D}\tilde{Q}\mbox{e}^{-F(-z_1,z_1,z_2,z_2)}
 &\sim& {\cal D}_1, \no\\
 \left<\str k\Lambda_{2}P^2\right>_{F^{(0)}(-z_1,z_1,z_2,z_2)}
 &=& -2\sum_{q\ne 0}\left(\Pi_+-\Pi_-^*\right), \no\\
 \int{\cal D}\tilde{Q}\mbox{e}^{-F(z_1,z_1,-z_2,z_2)}
 &\sim& {\cal D}_2, \no\\
 \left<\str k\Lambda_{1}P^2\right>_{F^{(0)}(z_1,z_1,-z_2,z_2)}
 &=& -2\sum_{q\ne 0}\left(\Pi_+-\Pi_-\right). \no\\
\ee
 For the last term in Eq.(\ref{W1AA}), we have  
\be
 \int{\cal D}\tilde{Q}\mbox{e}^{-F(-z_1,z_1,-z_2,z_2)}
 &\sim& {\cal D}_1{\cal D}_1{\cal D}_+^2{\cal D}_-^{-2}.
\ee
 The disconnected part is included in this contribution 
 and is subtracted to give the fourth and fifth terms in Eq.(\ref{rho21}).

 The purely connected part $W_2$ is calculated 
 using the asymptotic form of the Bessel function.
 We obtain 
\bw
\be
 W_2(\e_1,\e_2) &\sim& 
 \frac{1}{4\Delta^2}\int_1^\infty ds_1\int_0^1 ds_2 
 \biggl<\left\{\mbox{e}^{iz_1\left[s_1
 +\frac{1}{2}(s_1A_-+A_-^{(\Lambda)})\right]-\frac{i\pi}{4}}
 +(*) \right\} 
 \left\{\mbox{e}^{iz_1\left[s_2
 -\frac{1}{2}(s_2A_++A_+^{(\Lambda)})\right]-\frac{i\pi}{4}}
 +(*) \right\} \no\\
 & & \times 
 \left\{\mbox{e}^{iz_2\left[s_1
 +\frac{1}{2}(s_1A_--A_-^{(\Lambda)})\right]-\frac{i\pi}{4}}
 +(*) \right\} 
 \left\{\mbox{e}^{iz_2\left[s_2
 -\frac{1}{2}(s_2A_+-A_+^{(\Lambda)})\right]-\frac{i\pi}{4}}
 +(*) \right\}\biggr>_{\rm kin},
\ee
 where $(*)$ denotes the complex conjugate of the preceding term.
 Integrations of $s_{1,2}$ are evaluated 
 to find the asymptotic form
\be
 W_2(\e_1,\e_2) &\sim& 
 -\frac{1}{4\Delta^2}
 \int{\cal D}\tilde{Q}\biggl\{
 \frac{1}{(z_1-z_2)^2}\left[
 \mbox{e}^{-F(z_1,z_1,-z_2,-z_2)}
 -\mbox{e}^{2i(z_1-z_2)-F(-z_1,z_1,z_2,-z_2)}
 +(z_{1,2}\to-z_{1,2})\right] \no\\
 & & 
 +\frac{1}{(z_1+z_2)^2}\left[ 
 \mbox{e}^{-F(z_1,z_1,z_2,z_2)}
 +\mbox{e}^{2i(z_1+z_2)-F(-z_1,z_1,-z_2,z_2)}
 +(z_{1,2}\to-z_{1,2})\right] \no\\
 & & 
 +\frac{i}{z_1^2-z_2^2}\Bigl[
 \mbox{e}^{2iz_1-F(-z_1,z_1,z_2,z_2)}
 +\mbox{e}^{2iz_1-F(-z_1,z_1,-z_2,-z_2)} \no\\
 & & +\mbox{e}^{-2iz_2-F(z_1,z_1,z_2,-z_2)}
 +\mbox{e}^{-2iz_2-F(-z_1,-z_1,z_2,-z_2)}
 -(z_{1,2}\to-z_{1,2})\Bigr]
 \biggr\},
\ee
\ew
 Finally, keeping second order in $P$ for the functional $F$
 and using the formula (\ref{calD2}), 
 we obtain the result (\ref{rho22}). 

%%%%%%%%%%%%%%%%%%%%%%%%%%%%%%%%%%%%%%%%%%%%%%%%%%%%%%%%%%%%%%%%%%%%%%%%%%%%
%%%%%%%%%%%%%%%%%%%%%%%%%%%%%%%%%%%%%%%%%%%%%%%%%%%%%%%%%%%%%%%%%%%%%%%%%%%%

\end{document}